\documentclass[a4paper,11pt,onecolumn]{article}
\pdfoutput=1
\usepackage{jheppub}
\usepackage[T1]{fontenc}
\usepackage{graphicx}
\usepackage{dcolumn}
\usepackage{bm}
\usepackage{amsmath}
\usepackage{amssymb}
\usepackage{slashed}
\usepackage{accents}
\usepackage{subcaption}
\usepackage[percent]{overpic}
\usepackage{tikz}
\usepackage{float}
\def\d{d\kern-.8 ex\vrule height 1.3 ex depth-1.24 ex width .7 ex \kern .15 ex}
\def\D{D\kern-1.7 ex\vrule height .87 ex depth-.8 ex width .7 ex \kern .95 ex}

\newcommand{\be}{\begin{equation}}
\newcommand{\ee}{\end{equation}}
\newcommand{\bea}{\begin{eqnarray}}
\newcommand{\eea}{\end{eqnarray}}

\title{Random wave chaos on momentum-fractionated microstate geometries}

\author{Vladan \DJ uki\'c}
\author{and Mihailo \v{C}ubrovi\'c}
\affiliation{Center for the Study of Complex Systems, Institute of Physics Belgrade, University of Belgrade, Pregrevica 118, 11080 Belgrade, Serbia}
\emailAdd{djukic@ipb.ac.rs}
\emailAdd{mcubrovic@gmail.com}

\date{\today}

\abstract{

We study the role of momentum fractionation in the development of chaos on smooth 3-charge supergravity solutions of D1-D5 theory, specifically momentum-fractionated superstrata. We probe the chaos through the dynamics of null geodesics and waves. On fractionated backgrounds, both geodesics and waves show uniformly strong chaos (large chaotic sea and near-perfect agreement with the Berry random wave hypothesis). In contrast, geodesics and waves on single-mode (non-fractionated) superstrata are closer to separability (although still chaotic). We discuss a possibility that such dichotomy can be related to the expectation that monotone BPS states are less chaotic than fortuitous ones. Finally, we explain why our chaos probes are meaningful for microstate geometries even though they become trivial in the black hole limit.
}

\begin{document} 
\maketitle
\flushbottom

\section{Introduction} 

It is by now quite clear that the roads to black holes are paved with chaos. The state-of-the-art results on semiclassical horizons \cite{Sekino:2008he,Shenker:2013pqa,Shenker:2014cwa,Maldacena:2015waa,Cotler:2016fpe,Saad:2018bqo,Saad:2019lba} have established fast scrambling, maximum quantum Lyapunov exponent, and the linear ramp in the spectral form factor (SFF) as essential ingredients of semiclassical black holes. This is also seen from holography, on the CFT side, where maximum chaos is hard-wired into the workings of large-$N$, strongly coupled gauge theories. But we want more: the Holy Grail is to understand black holes in the stringy regime and to get rid of large $N$. Therefore, the puzzle of how to characterize chaos for black hole microstates in string theory is central for a number of research lines. One question is how to understand chaos in supersymmetric systems, which are, by definition, highly degenerate. Another is the microstate/fuzzball polemic \cite{Warner:2019jll,Bena:2022ldq,Raju:2018xue}: should black holes be understood as corresponding to exact, single solutions in string theory, or as averaged or smeared-out smooth solutions, which is also related to the puzzle of ensemble averaging in gravity in general \cite{Saad:2019lba,Pollack:2020gfa,Blommaert:2020seb,Blommaert:2021gha,Belin:2020hea,Belin:2023efa,DiUbaldo:2023qli,Belin:2026pko,Perlmutter:2026trb}.

The above questions have motivated us to explore the nature of chaos in supersymmetric microstate geometries and to compare it to what we know about black hole chaos \cite{Berenstein:2025ese,Djukic:2026ejk}. In particular, in \cite{Djukic:2026ejk} we have found a trend of increasing probe wave chaos across a family of smooth, horizonless supergravity backgrounds ordered by decreasing supersymmetry. But a puzzle remains: how to connect the trends of chaos to the question of black hole microstates. In other words, can the indicators of chaos provide a smoking-gun criterion on whether a state is a black hole microstate or not?

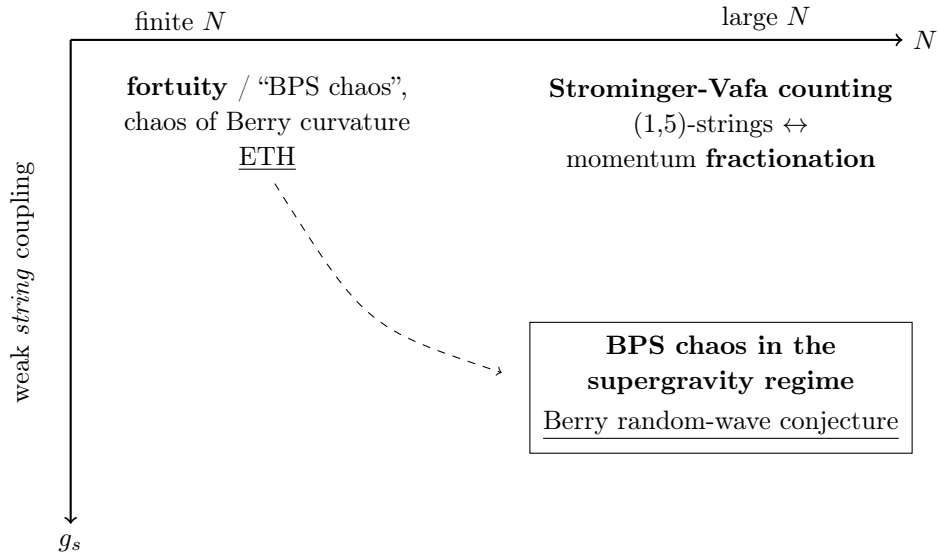
\begin{figure}[t]
\centering
\begin{tikzpicture}[x=1cm,y=1cm,font=\small]
  \draw[->,thick] (0,0) -- (11,0) node[right] {$N$};
  \draw[->,thick] (0,0) -- (0,-6.4) node[below] {$g_s$};
  \node[above left] at (2.2,0) {finite $N$};
  \node[above] at (9.2,0) {large $N$};
  \node[rotate=90,anchor=south] at (-0.35,-3.2) {weak \emph{string} coupling};
  \node[align=center] at (2.6,-1.1)
    {\textbf{fortuity} / ``BPS chaos'',\\ chaos of Berry curvature \\ \underline{ETH}};
  \node[align=center] at (8.6,-1.1)
    {\textbf{Strominger-Vafa counting}\\ (1,5)-strings $\leftrightarrow$\\ momentum \textbf{fractionation}};
  \node[draw,align=center,inner sep=5pt] at (8.6,-4.6)
    {\textbf{BPS chaos in the}\\ \textbf{supergravity regime}\\ [2pt]
     \underline{Berry random-wave conjecture}};
  \draw[->,dashed] (2.7,-1.9) .. controls (3.9,-3.8)  .. (5.7,-4.4);
\end{tikzpicture}
\caption{The criteria for black-holishness, and the regimes in which they are
computable, organized in the $(N,g_s)$ plane. Fortuity and the associated
``BPS chaos`` diagnostics are finite-$N$, weak-coupling statements; the counting that
they are meant to refine is a large-$N$, weak-coupling statement whose carriers
are the fractionally-moded (1,5)-strings. Microstate geometries -- and any
question about wave chaos on them -- live in the supergravity regime.}
\label{fig:map}
\end{figure}

The difficulty is that the criteria currently on offer and the regimes in which we can actually compute do not overlap. It is useful to organize them in the plane spanned by the charge $N \equiv N_1N_5$ and the string coupling constant $g_s$ (fig.~\ref{fig:map}). The upper-left corner -- weak coupling, finite $N$ --- is where the sharpest criteria live: fortuity and the BPS chaos diagnostics that accompany it \cite{Lin:2022zxd,Lin:2022rzw,Chen:2024oqv,Chen:2026vml}. The upper-right corner -- weak coupling, large $N$ -- is where we can actually \emph{count}: here lives the Strominger-Vafa result and the fractionated (1,5)-strings over which the $N_P$ units of momentum are distributed \cite{Strominger:1996sh, Bena:2013pda}. The lower-right corner -- the supergravity regime --- is where microstate geometries exist as classical solutions; this is where the chaos of probe waves can exist, and where our own diagnostics operate \cite{Djukic:2026ejk}. The logic of this paper is to travel along both axes: first horizontally, to argue that fortuity at finite $N$ and momentum fractionation at large $N$ are two faces of the same arithmetic for 3-charge systems; then vertically, to identify which supergravity solutions retain any track of that mechanism.

The recently proposed intriguing concept of fortuity \cite{Chang:2024zqi,Chang:2025rqy,Chang:2025wgo,Belin:2025hsg,Tierz:2026vrs,Belin:2026vfd,Berkooz:2026dbd} aims precisely at this same target -- identifying the black hole states -- but from the CFT perspective \cite{Kinney:2005ej,Grant:2008sk,Chang:2013fba}. The main lesson, established in \cite{Chang:2022mjp}, is that the spectrum is not exhausted by graviton-gas-like states. Following the terminology from \cite{Chang:2024zqi}, BPS cohomology classes are monotone if they persist as infinite sequences in $N$ (behaving essentially as a gas of (super)gravitons), and fortuitous if they exist only within a finite window of ranks $N$.

Fortuitous states therefore depend on $N$ in a highly erratic way and have no smooth $N \to \infty$ limit. In principle, fortuitous states might or might not be black hole microstates \cite{Chen:2025sum}, but black hole microstates likely have to be fortuitous. This is a sense in which fortuity is a \emph{finite}-$N$ criterion: it is a statement about how the spectrum jumps as $N$ is increased by $1$, and it is invisible in any strictly large-$N$ treatment. The same corner of fig.~\ref{fig:map} hosts the diagnostics that try to make this sharper -- BPS chaos \cite{Chen:2024oqv}, and the chaos of Berry curvature over the moduli space \cite{Chen:2026vml}. The idea of the ``BPS chaos'' approach is to cook up an indicator of chaos in the operator spectrum, inspired by the LMRS reasoning \cite{Lin:2022rzw,Lin:2022zxd}, which gets around the problem that BPS states exhibit high degeneracy and thus cannot show the textbook level repulsion. While this works at the formal level, the key puzzle is how to interpret the BPS chaos indicators in the only framework where the notion of ``quantum chaos'' has a rigorous meaning: in the regime where we can construct the \emph{semiclassical} limit, and show that this limit is chaotic.

\subsection{Idea 1: from fortuity to fractionation}

What is the statement analogous to fortuity for large $N$? This question can be made quantitative, in a form that speaks directly to the microstate geometry program \cite{Bena:2025pcy}. Consider first the $1/8$-BPS states obtained by exciting the left-moving sector of the D1-D5 CFT with the rigid generators like $L_{-1}$ alone -- precisely the generators out of which single-mode superstrata are assembled. These are in clean correspondence with the multi-particle supergraviton states around the bulk $\mathrm{AdS}_3 \times \mathbb{S}^3$ background, and in the Cardy regime $N_P \gg N$ they carry entropy $S_{\mathrm{supergravitons}}^{1/8-\mathrm{BPS}}\sim N^{1/2}N_P^{1/4}$ \cite{Shigemori:2019orj,Shigemori:2020yuo}. The most general $1/8$-BPS states are instead obtained by exciting arbitrary modes of the fundamental fields: on a strand of length $k$ these are the fractionally-moded oscillators, obtained by acting with \emph{affine} generators like $L_{-n/k}$. Counting the degrees of freedom then gives $S_{\mathrm{black~hole}}^{1/8-\mathrm{BPS}} \sim N^{1/2}N_P^{1/2} \gg S_{\mathrm{supergravitons}}^{1/8-\mathrm{BPS}}$. The entire gap between the two entropies, i.e., the whole difference between a gas of gravitons and a black hole, encapsulated in the factor $N_P^{1/4}$ -- is due to nothing but fractional moding. 

The two ends of the horizontal axis are now easy to glue together. The combinatorics of how the (1,5)-strings organize themselves into strands is arithmetically sensitive to $N_1$ and $N_5$ individually -- through their least common multiple $\mathfrak{lcm}(N_1,N_5)$, i.e., maximal strand length -- not merely to their product $N_1N_5$. At finite $N_1$ and $N_5$ this sensitivity is a genuine, erratic dependence on the charges: the available fractional modings, and with them the accessible states, jump as $\mathfrak{lcm}(N_1,N_5)$ changes. At large $N_1$ and $N_5$, as we know from elementary number theory, the smooth bilinear growth of $\mathfrak{lcm}$ dominates, the arithmetic fluctuations are subleading, and what survives is the smooth Strominger-Vafa envelope --- the regime in which the counting is a clean function of $N$ alone. In other words: \emph{momentum fractionation is what distinguishes black hole microstates from a gas of gravitons at large $N$, just like fortuity at finite $N$, and fortuity is what the arithmetic of fractionation looks like at finite $N$}. They are the same physics read at the opposite ends of the horizontal axis of fig.~\ref{fig:map}, and they share the same microscopic carriers -- the fractionally-moded (1,5)-strings.

Both statements above are weak-coupling statements: fortuity is formulated in the free symmetric orbifold CFT, and the counting is done on the D-brane open string description side. The chaos of wave propagation, however, is a question about \emph{geometry}, and geometries exist in the opposite corner of fig.~\ref{fig:map}. The well-known families of microstate geometries like 2-charge supertubes \cite{Lunin:2001fv,Lunin:2002iz}, 3-charge bubbling supertubes \cite{Giusto:2011fy,Giusto:2015dfa,Giusto:2012yz} and single-mode superstrata \cite{Bena:2015bea,Shigemori:2020yuo} are clearly monotone states. This important remark was made in \cite{Bena:2025pcy} and the argument is roughly that these geometries are controlled mainly by charges -- adding a unit of charge corresponds to changing $N_1$ or $N_5$ by 1, so a small change in $N_{1,5}$ gives a small change of geometry.

The present paper deals with \emph{momentum-fractionated} superstrata constructed in \cite{Bena:2016agb}. Unlike the single-mode $(k,m,n)$ superstrata whose momentum-carrying wave sits on a singly-wound supertube and is built from ordinary (integer) spectral flow, these solutions are built by adding momentum to a genuinely $k$-wound supertube via spectral interchange, so that the dual CFT operator involves \emph{fractional} spectral flow $J_{-1/k}^+$ -- equivalently, a filled Fermi sea of fractionally moded raising operators. This is a substantive generalization: it produces geometries whose throats are $k$ times deeper than those of a singly wound supertube, and whose dual CFT excitations are qualitatively more scrambled than anything reachable by ordinary spectral flow on unit length strands. The important question remains: \emph{do they represent typical black hole microstates?} Recall that in order to get full entropy one needs to act with a whole tower of fractionated modes $L_{-n/k}$, not just with $J_{-1/k}$, so fractionated superstrata we are dealing with here are still going to describe atypical black hole microstates, i.e. are not going to reproduce the full 3-charge black hole entropy estimated above.

Natural next candidates are \textit{bubbling~solutions} \cite{Bena:2005va,Bena:2006kb,Bena:2017fvm}, as outlined in \cite{Bena:2025pcy}, whose fortuitous nature is suggested by the fact that such geometries are not controlled by charges but by magnetic fluxes, and small changes of fluxes can still drastically change the solution or destroy it, as every flux depends on all charges, and in a complicated way. We leave the study of chaos for these backgrounds with diagnostics used in this work for the future.

\subsection{Idea 2: from ETH to Berry conjecture}

So far we have mainly discussed the puzzles of the string-theoretical description of black holes and how these could be uncovered through the study of chaos. But this line of research also has the other side: it is a challenge for chaos theory itself to formulate a meaningful notion of chaos in supersymmetric systems, with their high degeneracy and constraints that make it impossible to use the easiest textbook chaos criterion: level statistics. One way to do this is to perform the LMRS projection, as we have already mentioned in relation to \cite{Lin:2022rzw,Lin:2022zxd,Chen:2024oqv}. But, as we have explained, we want a criterion that can be extended all the way to the supergravity regime. Our diagnostic is not far in spirit from the eigenstate thermalization (ETH) ansatz for the matrix elements of some operator $\mathcal{O}$ sandwiched between the states with energies $E_i$ and $E_j$:
\be \label{eq:eth}
    \langle E_i | \mathcal O |E_j \rangle = f(E) \delta_{ij}+e^{-S(E)/2}R_{ij}, \quad E_i \approx E_j \equiv E.
\ee
The main point is the second term, which predicts that off-diagonal elements are effectively described by a Gaussian random matrix $R$. The ETH ansatz (\ref{eq:eth}) motivates the criteria of the upper-left corner of fig.~\ref{fig:map}. It is really the semiclassical avatar thereof. The point of ETH is that an individual eigenstate is statistically indistinguishable from a random vector drawn from the fixed-energy shell. The Berry random-wave conjecture \cite{Berry:1977} is the same assertion made in position space \cite{Srednicki:1994mfb}, for a system with chaotic classical limit: a high-lying eigenfunction on such a background behaves locally like a Gaussian random superposition of plane waves of fixed wavenumber $k_n$. This in turn allows one to compute the 2-point correlation function. In $\mathfrak D$ space dimensions the Berry random wave correlator reads
\be \label{eq:brw}
    C_{\mathfrak{D}}(\vec{x}_1,\vec{x}_2)\equiv\langle\Psi_n(\vec{x}_1)\Psi_n(\vec{x}_2)\rangle\propto J_{\frac{\mathfrak D-2}{2}}\left(k_n\left|\vec{x}_1-\vec{x}_2\right|\right).
\ee
Eqs.~(\ref{eq:eth}) and (\ref{eq:brw}) are presentations of one and the same idea in different bases -- matrix elements of a smooth operator in the first case, values of the wavefunction at points of configuration space in the other -- and a Gaussian random field of type (\ref{eq:brw}) is precisely what generates the $e^{-S/2}R_{ij}$ noise term in eq.~(\ref{eq:eth}) when a local operator is sandwiched between such, effectively random states. Finally, the ``BPS chaos'' of \cite{Lin:2022rzw,Lin:2022zxd,Chen:2024oqv} looks like a minimal modification of the ETH ansatz for matrix elements of operators with a BPS spectrum.

The punchline is that the random wave criterion comes from the same reasoning as the LMRS procedure, but in a formulation appropriate for the supergravity regime. In this regime the objects we have are a classical geometry and the wave operator of a probe field on it: its eigenfunctions are the only eigenstates available. However, the supergravity avatar comes with something which the CFT side lacks: the Berry conjecture is conditioned on ergodicity of the underlying \emph{classical} dynamics, so it carries a geometric prerequisite -- the absence of a (conformal) Killing tensor -- that can be checked directly on the metric. This is why the separability analysis that we perform in this paper is part of the same argument rather than a separate technical exercise, and that is why the criterion can be falsified by finding a hidden symmetry.

Another difference of our criterion from the BPS chaos is that the criteria of \cite{Chen:2024oqv,Chen:2026vml} are \emph{ensemble} statements: they ask for level statistics, or for the statistics of connection, over a family of BPS states within a fixed charge sector. Evaluating them therefore requires an explicit enumeration of that sector -- available only at small $N$ and low charge, or in sectors protected enough that the counting reduces to a clean generating function, or for a moduli space over which the states vary smoothly enough, so that a Berry connection can be defined and differentiated at all. Both requirements select exactly the states that are the least interesting for the black hole question: sectors whose BPS spectrum is under analytic control are generically the monotone ones, and a family smooth enough to carry a Berry connection is, by construction, a family defined uniformly in $N$. Fortuitous states resist both requirements at once. They exist only at isolated values of $N$ and are therefore not members of any smooth family. These diagnostics are thus structurally hardest to evaluate exactly where the physics they are meant to detect is claimed to live. 

On the other hand, random wave statistics on a given background are \emph{single-state} quantities. They require one explicitly specified coherent state -- not an ensemble, nor a smooth family of neighboring states. They are moreover evaluated at strong 't Hooft coupling, on the classical geometry, rather than at the free orbifold point.

The rest of the paper is organized as follows. In Section \ref{secintfrac} we describe the momentum-fractionated superstrata and study their separability properties analytically and numerically. In Section \ref{secwaves} we come to the core of the matter, studying the chaos of probe waves on different fractionated backgrounds and comparing it to the monotone microstates that we have studied in \cite{Djukic:2026ejk}. We will find differences in overall trends that can then be understood as manifestations of fractionation, as we argue in Section \ref{secfort}. Section \ref{secconc} sums up the conclusions.

\section{Integrability and Fractionated Superstrata}\label{secintfrac}

In this section we identify the supergravity solutions corresponding to the momentum-fractionated superstrata of \cite{Bena:2016agb} and then we study the integrability of geodesic motion in these backgrounds by inspecting the separability of the Hamilton-Jacobi equation for geodesics analytically and the structure of Poincar\' e sections numerically. We will conclude that the generic solution is nonintegrable, although there are special (zero measure) loci of integrable solutions.

\subsection{Generalities}

We work in six-dimensional $\mathcal N=1$ supergravity coupled to two anti-self-dual tensor multiplets, that arises from reducing type IIB string theory on $\mathbb{T}^4\times\mathbb{S}^1$ with circle radius $R_y$. As explained in \cite{Bena:2011dd}, the general BPS solution takes the form
\be \label{eq:met}
    ds^2_6 = - \frac{2}{\sqrt{\mathcal{P}}}(dv+\beta) \left [ du + \omega + \frac{1}{2}\mathcal F \left( dv + \beta \right)\right ] + \sqrt{\mathcal P} ds_4^2(\mathcal B),
\ee
where
\be
u\equiv\frac{t-y}{\sqrt{2}},\quad v \equiv\frac{t+y}{\sqrt{2}},
\ee
where $\mathcal P \equiv Z_1Z_2 - Z_4^2$ is built from three electric potentials $Z_1, Z_2, Z_4$ satisfying a linear first-layer BPS equation sourced by the base $\mathcal B$. The angular-momentum 1-form $\omega$ and the warp function $\mathcal F$ are determined by a second, quadratically-sourced layer \cite{Bena:2011dd,Giusto:2013rxa}. The starting point of \cite{Bena:2016agb} is the round, asymptotically flat, $k$-wound D1-D5 supertube known from \cite{Lunin:2001fv,Lunin:2001jy}: the base $\mathcal B$ is flat $\mathbb R^4$, written in bipolar coordinates $(r, \theta, \varphi_1, \varphi_2)$:
\be \label{eq:base}
    ds_4^2 (\mathcal B) = \Sigma \left( \frac{dr^2}{r^2+a^2} + d\theta^2 \right) + (r^2+a^2) \sin^2\theta d \varphi_1^2+r^2 \cos^2 \theta d\varphi_2^2, \quad \Sigma = r^2 +a^2 \cos^2 \theta,
\ee
dressed by the self-dual flux $\Theta_3 = d\beta$ of the dipole (winding) charge $k$. The result is a family of solutions whose CFT dual involves the fractionally moded current $J_{-1/k}^+$ rather than the ordinary integer spectral flow that builds the single-mode family.

Regularity of the resulting solution -- the absence of singular terms that a naive momentum-carrying deformation would otherwise produce -- is imposed by \textit{coiffuring}: an extra oscillating pair $(Z_4, \Theta_4)$ is introduced whose sole purpose is to cancel, algebraically, the singular pieces that a naive deformation of $(Z_1,Z_2)$ would produce, order by order in Fourier modes. There are two inequivalent coiffuring styles, labeled by integers $(k,p)$ \cite{Bena:2016agb}:
\begin{itemize}
    \item \textbf{Style 1:} both $Z_1$ and $Z_2$ oscillate at the same frequency as $Z_4$ ($n=p$), with two independent complex amplitudes $b_1,b_2$ fixed algebraically in terms of a single amplitude $b_4$;
    \item \textbf{Style 2:} $Z_2$ is left unexcited, $Z_1$ oscillates at twice the frequency of $Z_4$ ($n=2p$), and a single algebraic relation fixes $b_1$ in terms of $b_4$.
\end{itemize}
In both cases the physical momentum charge $Q_P$ is determined by the amplitudes and frequencies of oscillating modes as $Q_P \propto |b_4|^2/k^2R_y^2$, and the CFT dual carries genuine $1/k$-fractional moding, rather than the length-$1$ excitations of single-mode superstrata.

Before proceeding with calculations, we consider the dichotomy between the full solution for the metric (and other fields) and the decoupling limit, which zooms into the AdS region. Typical structure of the geometries from family (\ref{eq:met}-\ref{eq:base}) is the following: the outermost region, at largest $r$, is asymptotically flat, at smaller $r$ we enter the AdS${}_3$ throat, at still smaller $r$ we get a BTZ-like region, ending with a smooth cap at smallest $r$ scales. In order to make contact with holography and D1-D5 CFT states, and also with earlier work on chaos in smooth solutions \cite{Chen:2024oqv,Djukic:2026ejk}, we are interested mainly in the decoupling limit; it is also the easier case compared to the full geometry. For simple cases, it will turn out that we can say something about the separability in whole spacetime which then also holds in the decoupling limit; but most of the time we will focus exclusively on the decoupled AdS region. Our subsequent tests of wave chaos will also focus on the decoupling limit, whereas for geodesics we will sometimes also consider the motion in whole spacetime as the difference is negligible for geodesics (on the other hand, waves in the whole, asymptotically flat, spacetime would leak out and make the analysis much more difficult and incompatible with Berry random wave paradigm).

\subsection{Special Case: Coiffuring Style 1 with $Q_1=Q_2$}

The most tractable slice of the Style 1 family, the one we will work with throughout most of the paper, sets $Q_1=Q_2=Q$. We will show that this leads to cohomogeneity-2 metrics. Coiffuring forces $b_2=-b_1$, and the combination $b_1Q_1+b_2Q_2$ that would otherwise source the dependence of $\omega$ on the angular phase $\zeta$, defined as
\be
\zeta=\frac{v}{2R}-k\varphi_2,\label{eq:zeta}
\ee
vanishes identically. More strikingly, the warp factor $\mathcal P =Z_1 Z_2 -Z_4^2$ also loses all dependence on the phase even though $Z_1, Z_2, Z_4$ individually retain it, and since they never enter the metric except through $\mathcal P$, the full six-dimensional metric depends only on $(r,\theta)$: it is indeed cohomogeneity-2, with four commuting Killing vectors $\partial_u, \partial_v, \partial_{\varphi_1}, \partial_{\varphi_2}$ (rather than the expected three $\partial_u,\partial_{\varphi_1},\partial_{\zeta} = 2 R k \partial_v + \partial_{\varphi_2}$) that survive for generic $Q_1 \neq Q_2$. All the surviving $\zeta$-dependence is then carried entirely by the dilaton $e^{2\Phi}= Z_1^2/\mathcal P$ and the axion $C_0 = Z_4/Z_1$. In other words, a dilaton- or axion-coupled probe such as a string sees cohomogeneity-3 physics even at equal charges, while a minimal scalar does not.

The total 6-dimensional metric from eqs.~(\ref{eq:met}) and (\ref{eq:base}) becomes
\bea 
    ds^2_6 = &&- \frac{2}{\sqrt{\mathcal{P}}}(dv+\beta) \left [ du + \omega + \frac{1}{2}\mathcal F \left( dv + \beta \right)\right ] + \nonumber\\
    &&+~ \Sigma \sqrt{\mathcal P} \left( \frac{dr^2}{r^2+a^2} + d\theta^2 \right) + \sqrt{\mathcal P} \left[ (r^2+a^2) \sin^2\theta d \varphi_1^2+r^2 \cos^2 \theta d\varphi_2^2 \right], \label{eq:metsty2}
\eea
with $\omega = \omega_1 d\varphi_1 + \omega_2 d\varphi_2$, and
\bea
    \omega_1 &=& \frac{2 \sin^2 \theta}{\Sigma}\left [ kRa^2 + \frac{|b_4|^2}{kR} \frac{\left( r^2 + a^2 \right) \left( 1 - \Delta^{2kn} \right)}{\Lambda} \right], \label{eq:om1} \\
    \omega_2 &=& \frac{2 \cos^2 \theta}{\Sigma}\left [ kRa^2 + \frac{|b_4|^2}{kR} \frac{r^2\left( 1 - \Delta^{2kn} \right)}{\Lambda} \right], \label{eq:om2} \\
    \mathcal P &=& \left( 1 + \frac{Q}{\Sigma} \right)^2 - \frac{4 |b_4|^2 \Delta^{2kn}}{\Sigma^2}, \label{eq:P} \\
    \mathcal F &=& - \frac{|b_4|^2}{k^2 R^2 \Lambda} \left( 1 - \Delta^{2kn} \right), \label{eq:F} \\
    \beta &=& \frac{2k R a^2}{\Sigma} \left( \sin^2 \theta~d\varphi_1 - \cos^2 \theta~d\varphi_2 \right).\label{eq:beta}
\eea 
In the above we have also defined:
\be
\label{eq:lamdela}\Lambda=r^2+a^2\sin^2\theta,~~\Delta=\frac{a\cos\theta}{\sqrt{r^2+a^2}},~~a^2=\frac{Q^2-4\left|b_4\right|^2}{8k^2R^2}.
\ee
Notice that as $r \to \infty$ we get $\omega\to J\left(\sin^2\theta d\varphi_1+\cos^2\theta d\varphi_2\right)/r^2$ with $J=\left(Q^2+4|b_4|^2\right)/(4kR)$, reproducing the correct angular momentum. Setting $b_4=0$ collapses the solution (\ref{eq:om1}-\ref{eq:beta}) to the round, $k$-wound, asymptotically flat 2-charge D1-D5 supertube, from the Lunin-Mathur family \cite{Lunin:2001fv,Bena:2016agb}.

\subsubsection{(Non)separability of the Hamilton-Jacobi equation}

The resulting Style 1 $Q_1=Q_2$ metric (\ref{eq:metsty2}) has four exact, commuting Killing vectors, so the null-geodesic Hamiltonian admits five conserved quantities for free: the four momenta $p_u,p_v, p_{\varphi_1},p_{\varphi_2}$ conjugate to the cyclic coordinates, and the null constraint itself. Full Liouville integrability of the six-dimensional problem requires one further, generally hidden, conserved quantity, quadratic in momenta -- a (conformal) Killing tensor -- providing the analogue of Carter's fourth constant of motion in Kerr geometries. The natural diagnostic is to attempt the additive Hamilton-Jacobi ansatz
\be \label{eq:sepAnsatz}
    S=S_r(r)+S_\theta(\theta)+p_uu+p_vv+p_{\varphi_1}\varphi_1+p_{\varphi_2}\varphi_2
\ee
and check whether it separates the Hamilton-Jacobi equation
\be \label{eq:hjgen}
    g^{MN}\partial_MS\partial_NS=0,\quad M,N\in\{u,v,r,\theta,\varphi_1,\varphi_2\}.
\ee
The separability analysis is given in Appendix \ref{secappa}. It encompasses a number of variable transformations and a detailed analysis of the algebraic structure of the effective potential seen by the null geodesic; while every single step is elementary, altogether the derivation is rather long, so we have decided to leave it out of the main text. When the dust settles, we find that the Hamilton-Jacobi equation (\ref{eq:hjgen}) acquires the form
\be
\left(r^2+a^2\right)\left(\partial_rS_r\right)^2+\left(\partial_\theta S_\theta\right)^2=\Sigma\mathcal W,
\ee
which is manifestly separable, except for the term $\Sigma\mathcal{W}$. The question thus boils down to the structure of $\Sigma\mathcal{W}$. After the transformations described in the Appendix, that term can be written as the sum of the $b_4=0$ part and the remainder:
\be
    \Sigma \mathcal W = \Sigma \mathcal W \big |_{b_4 =0} - \frac{1}{2}\mathcal C \mathcal F, \quad \mathcal C \equiv 8 k R p_u (m_1 +m_2 + 2 k R p_v).\label{eq:potfinal}
\ee
But the $b_4=0$ case reduces to the well-known round supertube, proven to be separable in \cite{Chervonyi:2013eja}. Therefore, analyzing (\ref{eq:potfinal}) we get four cases:
\begin{enumerate}
\item $b_4=0$. From (\ref{eq:F}) this leads to $\mathcal F=0$. The potential (\ref{eq:potfinal}) reduces to just the round supertube contribution, so it is separable.
\item $kn=1$. From (\ref{eq:lamdela}) this implies $1-\Delta^2=\frac{r^2+a^2-a^2\cos^2\theta}{r^2+a^2}=\frac{\Lambda}{r^2+a^2}$ therefore, from (\ref{eq:F}), $\mathcal F\propto\frac{1-\Delta^2}{\Lambda}=\frac{1}{r^2+a^2}$, which is a function of $r$ alone and does not spoil the separability of the first term. We thus have another separable case.
\item In the special loci defined by $p_u=0$ or $m_1+m_2+2kRp_v=0$ (for any $b_4$ and $kn$), we get $\mathcal{C}=0$ from (\ref{eq:potfinal}). Therefore, the potential reduces to the first term, again becoming separable.
\item The remaining cases correspond to $b_4 \neq 0$ and $kn\geq 2$, outside the special locus from the previous case. They yield nonseparable motion.
\end{enumerate}

At this place we have to warn of a subtlety in the above analysis which is easy to get wrong. While the \emph{existence} of a Killing tensor is a coordinate-independent fact, the question whether a given specific additive ansatz \emph{solves the equation} is not: it depends on whether the chosen coordinates happen to be adapted to the tensor's eigenvector structure. This phenomenon was observed in the black hole literature: the Boyer-Lindquist-type coordinates, which yield a separable Hamilton-Jacobi equation for geodesics in the Kerr-Newman black hole background, fail to do so for generic higher-dimensional rotating black holes. This initially \emph{``misled people to believe that the geodesic motion in these spacetimes is no longer integrable''} (quoted from \cite{Frolov:2017kze}), but eventually the discovery of the principal conformal Killing-Yano tensor revealed the correctly adapted (``canonical'') coordinates and established integrability. We therefore have to treat the failure of the additive ansatz in any coordinates as evidence to interpret with care, not as a proof of nonintegrability.

We can go one step further. Motivated by the findings of \cite{Bena:2020yii,Djukic:2026ejk} we consider the deep-throat limit. The throat length can be expressed as a dimensionless ratio of the AdS scale $\sim\sqrt{Q_1Q_2}$ and the position of the cap which is roughly $r\sim a$. We can thus use $D\equiv\sqrt{Q_1Q_2}/a$ as the proxy for the throat length ($\log D$ is then the redshift factor). The long-throat regime is then taken by fixing $Q$ and taking $a\to 0$, which gives $D\to\infty$. From (\ref{eq:lamdela}) this also means $\Delta\to 0$ and this, according to (\ref{eq:F}), implies that the function $\mathcal F$ depends on the coordinates only through $\Lambda$, which has a simple form (eq.~(\ref{eq:lamdela})). Expanding over the small ratio $a/r$ we get:
\be \label{eq:deepthroatlimit}
\mathcal F =-\frac{|b_4|^2}{k^2 R^2 r^2}\left[1-\frac{a^2\sin^2\theta}{r^2}-\left(\frac{a^2\cos^2 \theta}{r^2+a^2}\right)^{kn}+\mathcal O \left(\frac{a^4}{r^4}\right)\right].
\ee
Crucially, the leading term in the throat $r > a$ is a function of $r$ alone, hence the Hamilton-Jacobi equation becomes separable in the deep-throat limit, \emph{for any parameter values}. This reproduces the known emergent conformal symmetry of black hole throats \cite{Frolov:2017kze,Castro:2010fd}. Therefore, geodesics and waves that live (mostly) inside the throat will show regular motion in the infinite throat limit. Note, however, that the order of limits is important: the above argument does not give separability in the cap $r\lesssim a$, therefore the probes that live mostly inside the cap will never become regular, no matter how deep the throat is. 

\subsubsection{Numerical evidence: Poincar\'e sections of the null-geodesic flow}

We have found that the additive ansatz (\ref{eq:sepAnsatz}) does not work for the case (4) of the list given in the previous section. But, as we have already mentioned, this is a statement about one coordinate system, and its failure proves nothing on its own. A Poincar\' e section is not subject to that caveat: invariant curves on a surface of section are a coordinate-independent signal of a second commuting constant of motion on that energy shell, and their destruction is a coordinate-independent signal of its absence. We therefore integrate the null geodesics of (\ref{eq:metsty2}) numerically and take that as strong evidence of the nonintegrability, while the above findings for the Hamilton-Jacobi equation serve as a useful analytic insight but not a proof.\footnote{On the other hand, the cases (1-3) where we have found separability do constitute a proof. We know that these cases are separable but do not know for sure that the case (4) is nonseparable before performing numerical checks.}

The whole problem collapses to a 2-degree-of-freedom system with a single potential. On null geodesics we have the constraint $H=0$, and the orbits can be found from the rescaled Hamiltonian
\be \label{eq:ham}
    \widetilde H = \frac{1}{2} \left[ \left( r^2 + a^2 \right) p_r^2 + p_\theta ^2 - \Sigma \mathcal W(r,\theta) \right] \quad \mathrm{on} \quad \widetilde H = 0,
\ee
which differs from $H$ only by the rescaling of the affine parameter by $\sqrt{\mathcal P} \Sigma$. The classically allowed region is $\mathcal W \geq 0$.

Two remarks on the calculation are in order. First, a surface of section must lie on a single level set of all the conserved quantities: we fix $\{ p_u, p_v,m_1, m_2 \}$ and impose $\widetilde H =0$ by solving for $p_\theta >0$ at each seed, so that every orbit lies on the same 3-dimensional shell. Second, $p_u p_v = p_t^2 - p_y^2$ controls whether the motion is bound; trapped null geodesics require a forbidden band in $r$, which for the asymptotically flat metric only exists over a finite range of $m_1/Q,~m_2/Q$.

\begin{figure}[H]
\centering
(a)\includegraphics[width=.45\linewidth]{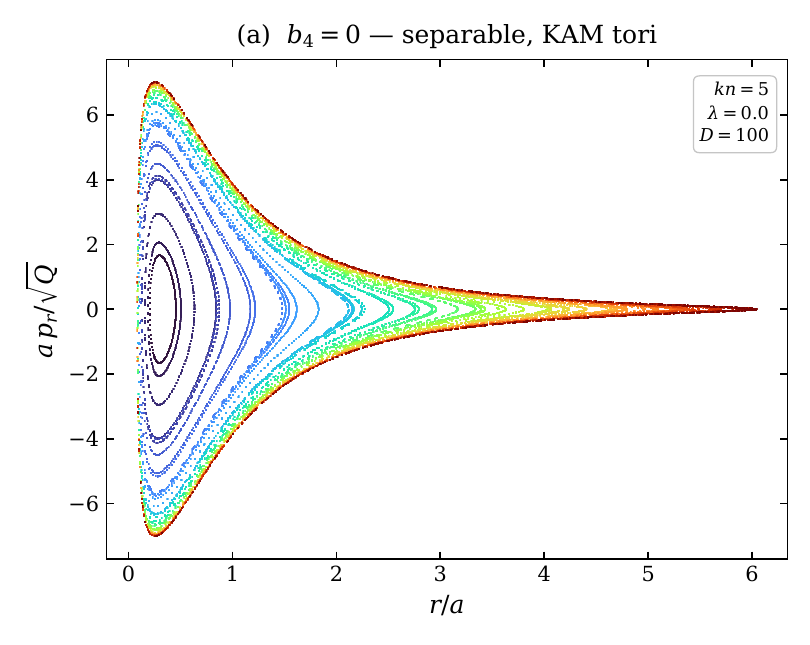}
(b)\includegraphics[width=.45\linewidth]{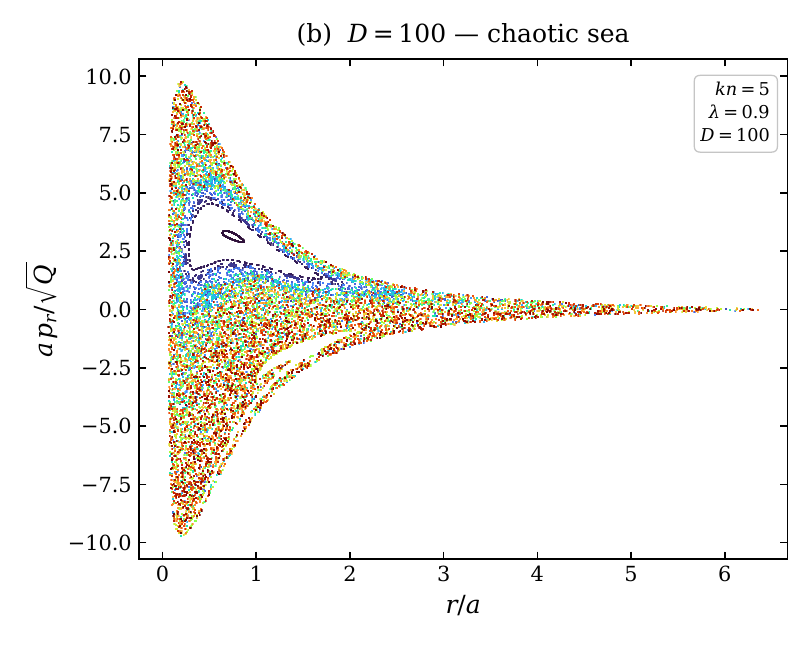}
(c)\includegraphics[width=.45\linewidth]{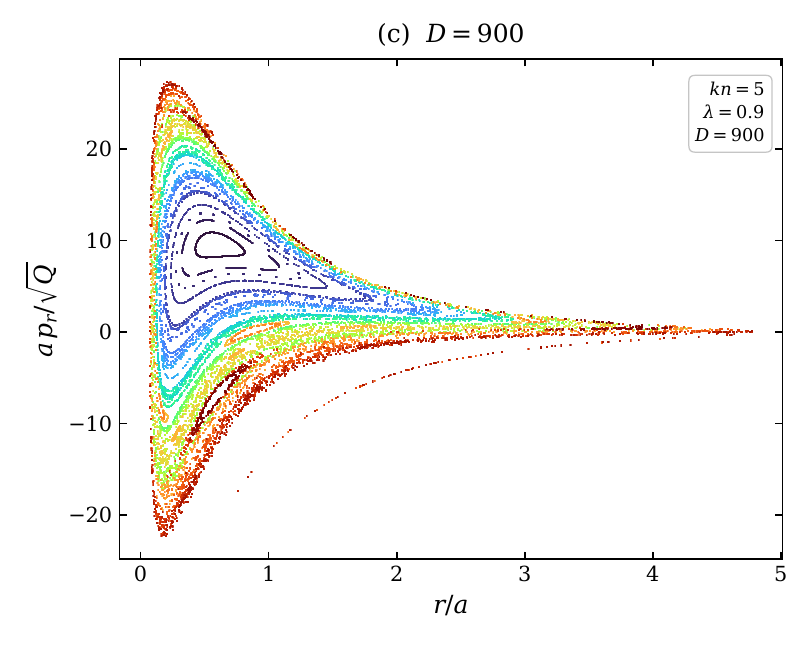}
(d)\includegraphics[width=.45\linewidth]{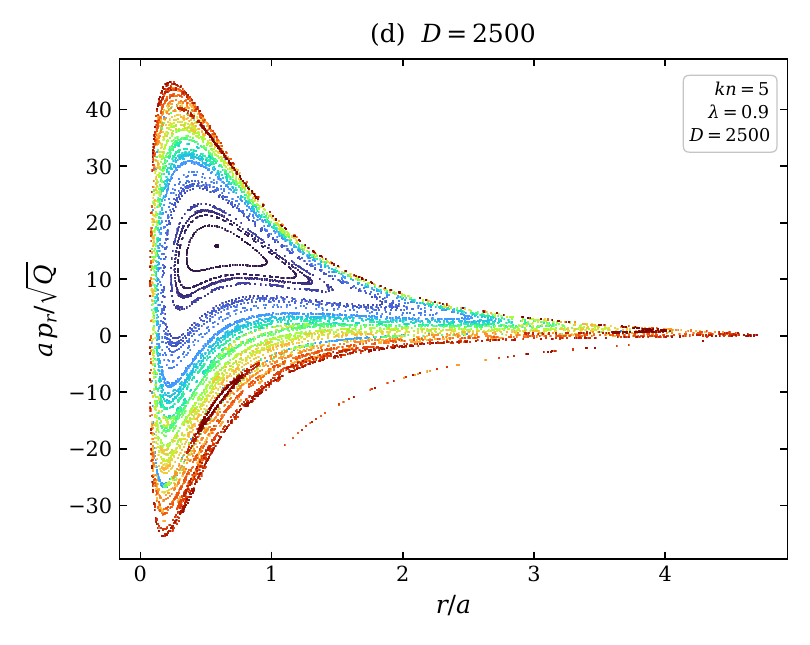}
\caption{Poincar\' e sections for null-geodesic flow on the Style 1, $Q_1=Q_2\equiv Q$ metric (\ref{eq:metsty2}), at the surface of section $\theta = \pi/4, p_\theta >0$. (a) $b_4 = 0$: the separable 2-charge supertube, whose shell is filled by invariant curves. (b) $\lambda = 2 |b_4|/Q = 0.9$ at throat depth $D=Q/a=100$: chaotic sea with one surviving regular island. (c, d) the same deformation and the same dimensionless charges at $D=900$ and $D=2500$; the sea recedes and invariant curves multiply again. All panels are asymptotically flat with $k=1, n=5$, $p_u = p_v = 1/2$, $\widetilde m_1 \equiv m_1/Q = 0.1, \widetilde m_2 \equiv m_2/Q = -0.067$, and all orbits share the same values of $(p_u,p_v,\widetilde m_1, \widetilde m_2)$.}
\label{fig:momntfracssPS}
\end{figure}

At $b_4 = 0$, where we have proven separability, we expect a foliation of every energy shell by invariant tori, and this is what we find. Fig.~\ref{fig:momntfracssPS}(a) shows the section $\theta = \pi/4, p_\theta >0$ of the 2-charge supertube: the shell is filled by smooth nested closed curves down to the resolution of the integration, with no islands and no visible stochastic layer. This is a genuine, if unsurprising, check on the analytic separability result.

\begin{figure}[H]
\centering
(a)\includegraphics[width=.45\linewidth]{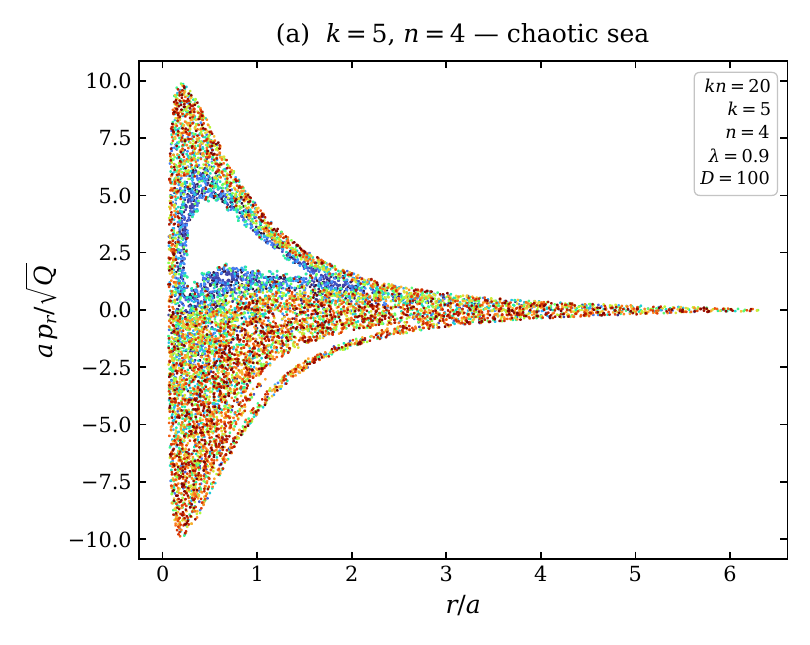}
(b)\includegraphics[width=.45\linewidth]{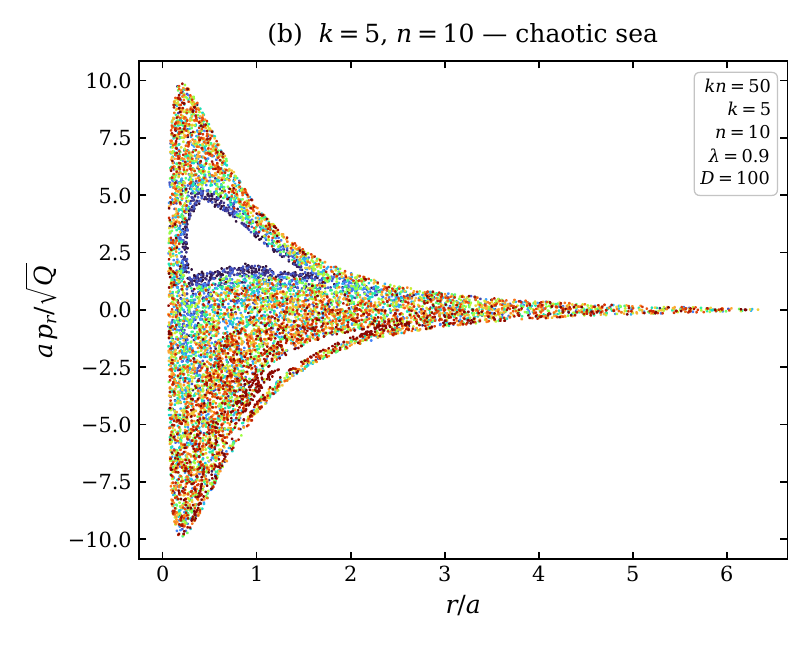}
(c)\includegraphics[width=.45\linewidth]{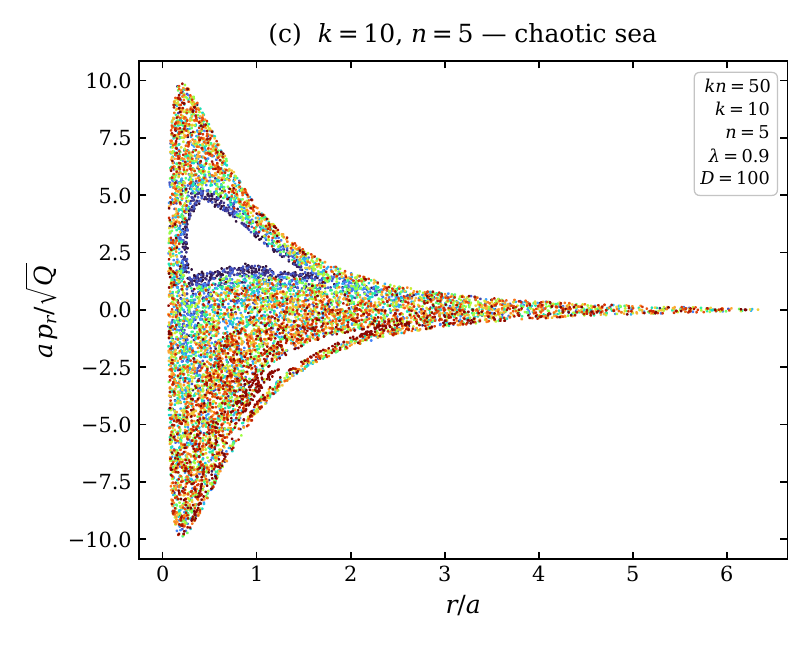}
(d)\includegraphics[width=.45\linewidth]{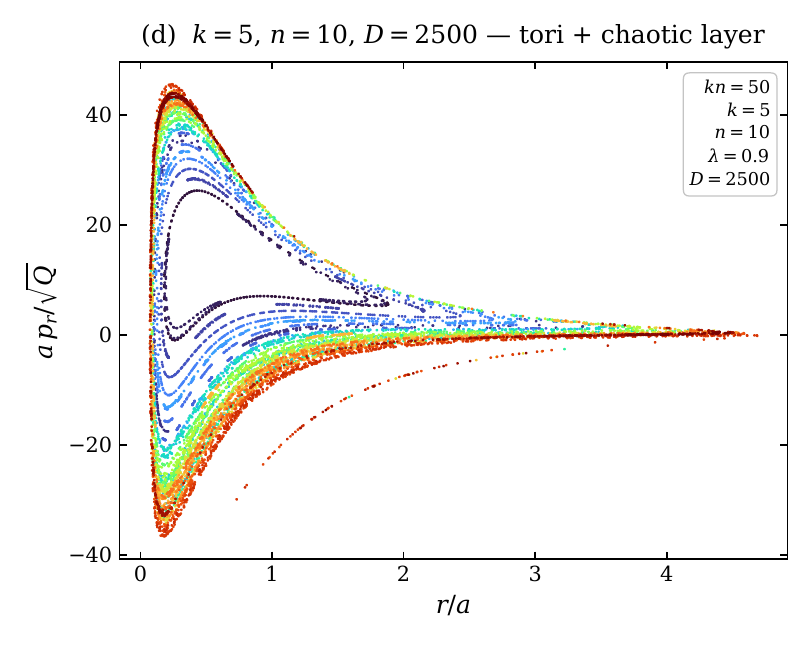}
\caption{Poincar\' e sections for null-geodesic flow on the Style 1, $Q_1=Q_2=Q$ metric (\ref{eq:metsty2}), at the surface of section $\theta = \pi/4, p_\theta >0$. (a) $k=5,~n=4~(kn=20)$ at throat depth $D=Q/a = 100$: the shell is filled by chaotic sea. (b) $k=5,~n=10~(kn=50)$, same $\lambda$ and $D$: raising $n$ (and hence $kn$) does not visibly enlarge the chaotic sea. (c) $k=10,~n=5$ and hence the same product $kn=50$ as in (b) with a different value of $k$: the section is almost identical to (b). (d) $k=5,~n=10$ at $D=2500$: deepening the AdS${}_2$ throat pushes the sea back and invariant curves refill most of the shell, but the phase space remains mixed rather than completely integrable -- a thin stochastic layer survives. All panels are asymptotically flat with $p_u = p_v = 1/2$, $\widetilde m_1 \equiv m_1/Q = 0.1, \widetilde m_2 \equiv m_2/Q = -0.067$, and all orbits within a panel share the same values of $(p_u,p_v,\widetilde m_1, \widetilde m_2)$.}
\label{fig:momntfracssPS2}
\end{figure}

For $b_4\neq 0$, switching on the momentum wave with $kn \geq 2$ is expected to destroy integrability. It is convenient to use a scaling symmetry (manifest in eq.~(\ref{eq:hj}) in the Appendix) that transforms $r \to \tilde \mu r, a \to \tilde\mu a, Q \to \tilde\mu^2 Q, b_4 \to \tilde \mu^2 b_4, k R \to \tilde\mu kR$ to set $a=1$, so that the independent dimensionless data are
\be \label{eq:lambdadef}
    \lambda = \frac{2 |b_4|}{Q} \in [0,1), \quad D \equiv Q/a, \quad kn, \quad p_u,~p_v,~ \widetilde m_1 \equiv m_1/Q,~\widetilde m_2 \equiv m_2/Q.
\ee
Notice that $\lambda$ is not an independent tuning knob but is related to the momentum charge: from $Q_P = |b_4|^2/2 k^2 R^2$ and $a^2 = (Q^2 - 4 |b_4|^2)/8 k^2 R^2$ one finds that $Q_P k^2 R_y^2/Q_1Q_2 = \lambda^2$, so that $\lambda\to 1$ is exactly the scaling limit $Q_P \to Q_1Q_2 / k^2 R_y^2$, with $a \to 0$.

Fig.~\ref{fig:momntfracssPS}(b) shows the same shell as \ref{fig:momntfracssPS}(a) at $\lambda =0.9$ with $kn=5$: most of it has transformed into a single connected chaotic sea, with one surviving island of regular motion around a stable periodic orbit. Panels (c) and (d) hold the deformation and the dimensionless charges fixed but lengthen the throat: $D = 100 \to 900 \to 2500$ (b, c, d respectively). As the throat becomes very long, the chaotic sea shrinks and invariant curves refill the shell, as expected from the discussion around eq.~(\ref{eq:deepthroatlimit}). A similar trend is observed in a sequence of Poincar\' e sections displayed in fig.~\ref{fig:momntfracssPS2}, where we show that the crucial parameter controlling the chaotic sea is the throat length $D$, not other parameters individually (we will discuss the role of $D$ in detail later). Also, we again see the geodesic chaos shrink as the throat becomes very long.

\subsection{General Coiffuring Style 1 Family: $Q_1\neq Q_2$}
\label{sec:gencoiff}

We will now go back to the general class of supergravity solutions arising from coiffuring via Style 1 \cite{Bena:2016agb}. In order not to have singularities in the solution, coiffuring imposes the conditions $b_4^2 = Q_1Q_2 b_1 b_2$ and $(b_1+b_2)Q_1Q_2 + a^2 (b_1Q_1+b_2 Q_2) = 0$, which is solved by
\be
    b_1 = \frac{ib_4}{Q_1} \chi, \quad b_2 = -\frac{i b_4}{Q_2 \chi}, \qquad \chi \equiv \sqrt{\frac{Q_1+a^2}{Q_2+a^2}}.
\ee
The second layer of equations then yields
\bea
    &&\omega_r = -\frac{i}{4 kR}\frac{r \Delta^{kn}}{r^2+a^2} \left( \mathcal C_1 e^{-i n \zeta}- \bar{\mathcal C}_1 e^{in \zeta} \right), \quad \omega_\theta =\frac{i \Delta^{kn}}{4kR}\tan \theta \left( \mathcal C_1 e^{-i n \zeta}- \bar{\mathcal C}_1 e^{in \zeta}  \right),~~~~\label{eq:qgenomegar} \\
    && \omega_1 = - \frac{a^2}{4kR}\frac{\Delta^{kn}\sin^2\theta}{\Sigma} \left( \mathcal C_1 e^{-i n \zeta}+ \bar{\mathcal C}_1 e^{in \zeta} \right)  \nonumber \\
    && ~~~~~~~ -\frac{2 |b_4|^2}{kR} \frac{r^2+a^2}{\Sigma} \left[ \frac{\left( \Delta^{2kn} -1\right) \sin^2\theta}{\Lambda} + \frac{1}{a^2} \right]+ \frac{J}{a^2} \frac{r^2+a^2}{\Sigma} + c_1, \label{eq:om1gen} \\
    && \omega_2 = \frac{r^2}{4kR}\frac{\Delta^{kn}}{\Sigma} \left( \mathcal C_1 e^{-i n \zeta}+ \bar{\mathcal C}_1 e^{in \zeta} \right) \nonumber \\
    && ~~~~~~~ - \frac{2 |b_4|^2}{kR} \frac{r^2}{\Sigma} \left[ \frac{\left( \Delta^{2kn} -1\right) \cos^2\theta}{\Lambda} - \frac{1}{a^2} \right] - \frac{J}{a^2} \frac{r^2}{\Sigma} + c_2, \label{eq:om2gen} \\
    && \mathcal F = - \frac{|b_4|^2}{k^2R^2}\frac{1- \Delta^{2kn}}{\Lambda}. \label{eq:Fgen}
\eea
 
The remaining functions from the above equations are defined as
\bea
&&\mathcal C_1\equiv b_1Q_1+b_2Q_2=ib_4 \left(\chi-\chi^{-1}\right)\\
&&c_1=-2kR,~~c_2=2kR,~~J=\frac{Q_1Q_2+4|b_4|^2}{4kR},~~a^2=\frac{Q_1Q_2-4|b_4|^2}{8k^2R^2}\\
&&\beta=\frac{2kRa^2}{\Sigma}\left(\sin^2\theta d\varphi_1-\cos^2\theta d\varphi_2\right).
\eea
Constructing $\mathcal{P}$ as $\mathcal P =Z_1Z_2-Z_4^2$, exactly one harmonic appears:
\be \label{eq:Pgen}
\mathcal P = \left(1+\frac{Q_1}{\Sigma}\right)\left(1+\frac{Q_2}{\Sigma}\right)-\frac{4|b_4|^2 \Delta^{2kn}}{\Sigma^2}+2\Re\left[\mathcal C_1 \frac{\Delta^{kn}}{\Sigma}\left(1-\frac{a^2}{\Sigma}\right)e^{-in\zeta}\right].
\ee
Now that the metric and the relations among the parameters are specified, the stage is set for the study of separability. The technical details we again leave out from the main text; they can be found in Appendix \ref{secappb}. In a nutshell, from the conservation laws and the algebraic structure of the geodesic action one finds that the null geodesic motion can be packaged into an effective Hamiltonian $\bar{H}$ with 3 degrees of freedom:
\be \label{eq:hamcoh3}
\bar{H}=\left(r^2+a^2\right)\left(p_r-\omega_rp_u\right)^2+\left(p_\theta-\omega_\theta p_u\right)^2-A_2m^2-A_1m-A_0=0.
\ee
This Hamiltonian is not so difficult to analyze. Making the ansatz
\be\label{eq:ansatz2}
S=p_uu+p_vv+m_1\varphi_1+m_2\varphi_2+S_r(r)+S_{\theta}(\theta)+S_\zeta(\zeta)
\ee
and using (\ref{eq:hamcoh3}) we get the Hamilton-Jacobi equation
\be \label{eq:hjcoh3}
    \left(r^2 +a^2 \right) \left(\partial_rS_r - \omega_r p_u\right)^2 + \left(\partial_\theta S_\theta-\omega_\theta p_u \right)^2 = \Sigma \mathcal W,
\ee
where we have dumped the terms proportional to $A_0$, $A_1$, $A_2$ into the potential term $\mathcal{W}$, rescaled by $\Sigma$ for convenience. Explicit expressions for $A_0$, $A_1$, $A_2$ are given in eqs.~(\ref{eq:a2}-\ref{eq:a0}). There are two independent obstructions to separability. First, $\omega_r$ depends on $\theta$ and $\omega_\theta$ on $r$, both through $\Delta^{kn}$, so the left-hand side is not of the form $f(r) + g(\theta)$ even at fixed $\zeta$. Second, the right-hand side must be $\zeta$-independent, which in turn requires $\partial_{\zeta}(A_2m^2+A_1m+A_0) = 0$ and returns us to $Q_1 = Q_2$. Let us show this by first isolating $\zeta$-dependent from $\zeta$-independent terms in the functions that appear in the metric. We introduce
\be
\sigma = \sin \left(n \zeta \right)\label{eq:sigma}
\ee
and rewrite the metric functions as:
\bea
    \text{eq.~(\ref{eq:Pgen}) becomes} &&\quad \mathcal P = \mathcal P_0 + 2 |\mathcal C_1| \mathfrak p \sigma, ~\mathfrak p \equiv \frac{\Delta^{kn}}{\Sigma} \left( 1- \frac{a^2}{\Sigma} \right) \\
    \text{eqs.~(\ref{eq:om1gen}) and (\ref{eq:om2gen}) become} && \quad \omega_i = \omega_i^0 + |\mathcal C_1|\mathfrak m_i \sigma
\eea
where $\mathfrak{m}_1 = - \frac{a^2 \Delta^{kn}\sin^2 \theta}{2kR \Sigma},~\mathfrak m_2 = \frac{r^2 \Delta^{kn}}{2kR \Sigma}$, while $\mathcal F$ and $\beta$ are $\zeta$-independent. We now define
\be
\mathcal M_i = \mathcal M_i^0 - |\mathcal C_1|\mathfrak{m}_i p_u \sigma,
\ee
where $\mathcal{M}_i^0$ is the $\zeta$-independent part (full definition of $\mathcal{M}_i^0$ and $\mathcal{M}_i$ is found in Appendix \ref{secappb}). Now, according to eqs.~(\ref{eq:a2}-\ref{eq:a0}), $A_2$ is independent of $\zeta$, while $A_1$ and $A_0$ depend on $\zeta$ through $\mathcal{M}_i$, which in turn depend on $\sigma$ (given in eq.~(\ref{eq:sigma})). Also, $A_1$ is linear and $A_0$ quadratic in $\mathcal M_i$, so we get that
\be
A_1m + A_0 = X_0(r,\theta)+ X_1(r,\theta)\sigma + X_2(r,\theta)\sigma^2
\ee
is a \emph{quadratic} polynomial in $\sigma$ -- no higher harmonics appear. Therefore,
\be
    \partial_\zeta \left( A_1 m + A_0 \right) = n \cos \left( n \zeta \right) \left[ X_1 + X_2 \sigma \right],
\ee
which should vanish for all $\zeta$. This in turn requires $X_1=X_2 = 0$ separately. One can easily show that both have the following simple form ($\Theta_i$ are some functions of their arguments):
\be
    X_i = |\mathcal C_1|p_u \Theta_i (r,\theta; m, m_1, m_2, p_v),
\ee
provided that the following condition is satisfied:
\be
    \partial_\zeta \left( A_1 m + A_0 \right) = 0 \quad \Rightarrow \quad  \mathcal C_1 = 0 ~\text{or}~ p_u = 0.
\ee
The $p_u=0$ locus is the separable locus that we have already encountered and we will comment on it below. For the first possibility ($\mathcal{C}_1=0$) we have $\mathcal C_1 = i b_4 \left( \chi - \chi^{-1} \right) = 0$ with $\chi = \sqrt{(Q_1 + a^2)/(Q_2+a^2)}$, so for $b_4 \neq 0$ we need $Q_1=Q_2$. The additive ansatz therefore fails for this family whenever $Q_1 \neq Q_2$.

Nonetheless, both terms that obstruct separability carry a factor of $p_u$: the minimal coupling term in eq.~(\ref{eq:hjcoh3}) manifestly so, and the $\zeta$-dependence because $\mathcal P$ enters only through $2p_u p_v^{(m)}$ and $\mathcal F p_u^2$, while $\omega_r,~\omega_\theta,~\omega_1,~\omega_2$ enter only as $\omega p_u$. At $p_u = 0$,
\be
    \Sigma \mathcal W|_{p_u = 0} = - \Sigma \left[ \frac{\left( m_1 - \beta_1 p_v^{(m)} \right)^2}{\left( r^2 + a^2\right) \sin^2 \theta} + \frac{\left( m_2^{(m)} - \beta_2 p_v^{(m)} \right)^2}{r^2 \cos^2 \theta} \right],
\ee
which contains neither $\mathcal P$, nor $\mathcal F$, nor any $\omega$. Hence the system collapses to two degrees of freedom, and $\partial_r\partial_\theta (\Sigma \mathcal W)|_{p_u = 0} = 0$ exactly. Therefore, $p_u = 0$ is an exact separable locus. There is also the equatorial separable locus at $\theta = \pi/2$, since all $\zeta$-couplings carry the $\Delta^{kn}$ factor, which vanishes at $\theta = \pi/2$.

In conclusion, except for the trivial case when it degenerates to the $Q_1=Q_2$ system, the general coiffuring Style 1 family seems to be separable only on the zero-measure loci $p_u=0$ and $\theta=\pi/2$. We will now support this conclusion by numerics.

\begin{figure}[ht]
\centering
(a)\includegraphics[width=.29\linewidth]{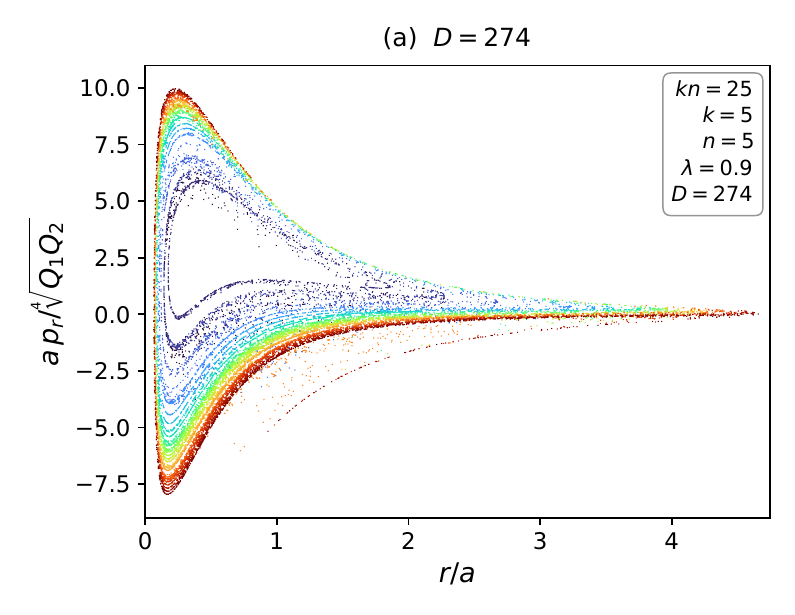}
(b)\includegraphics[width=.29\linewidth]{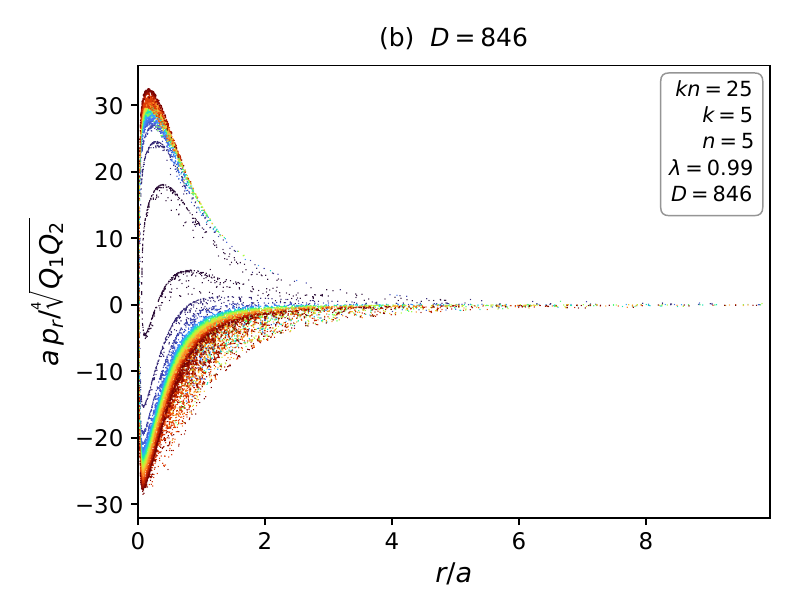}
(c)\includegraphics[width=.29\linewidth]{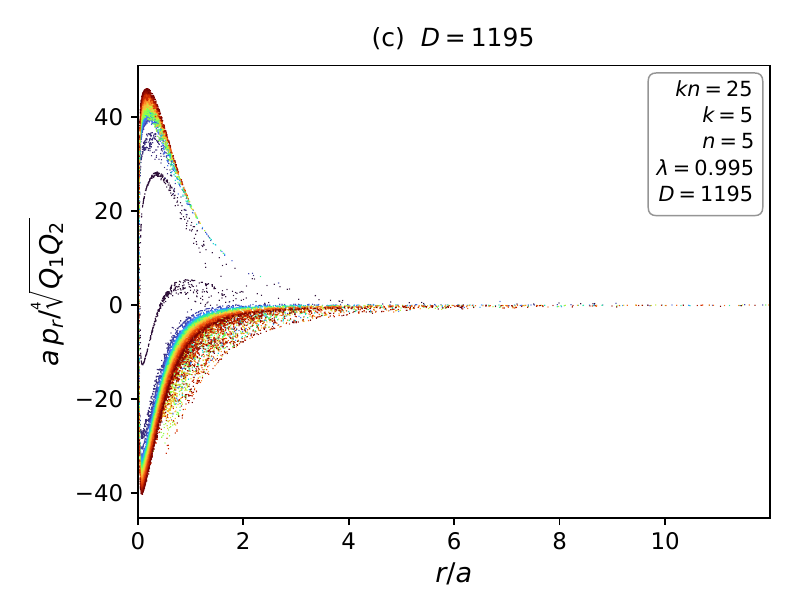}
(d)\includegraphics[width=.29\linewidth]{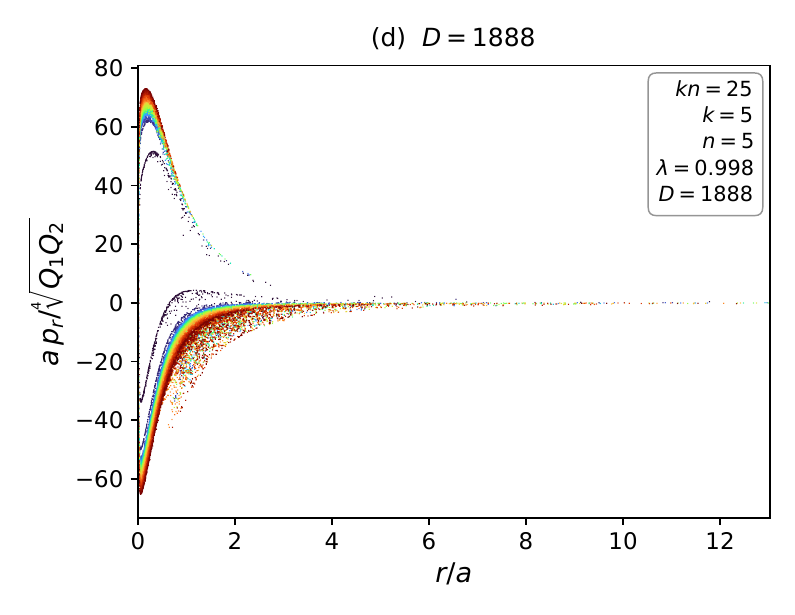}
(e)\includegraphics[width=.29\linewidth]{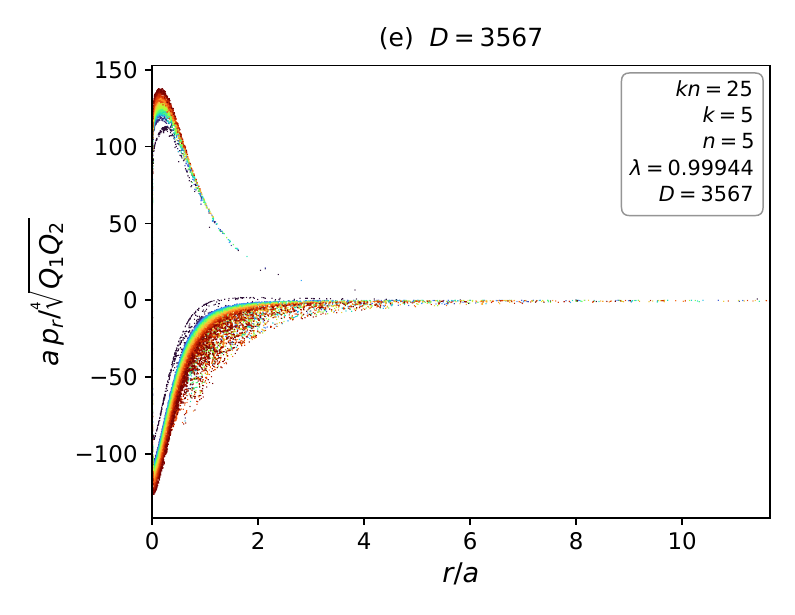}
(f)\includegraphics[width=.29\linewidth]{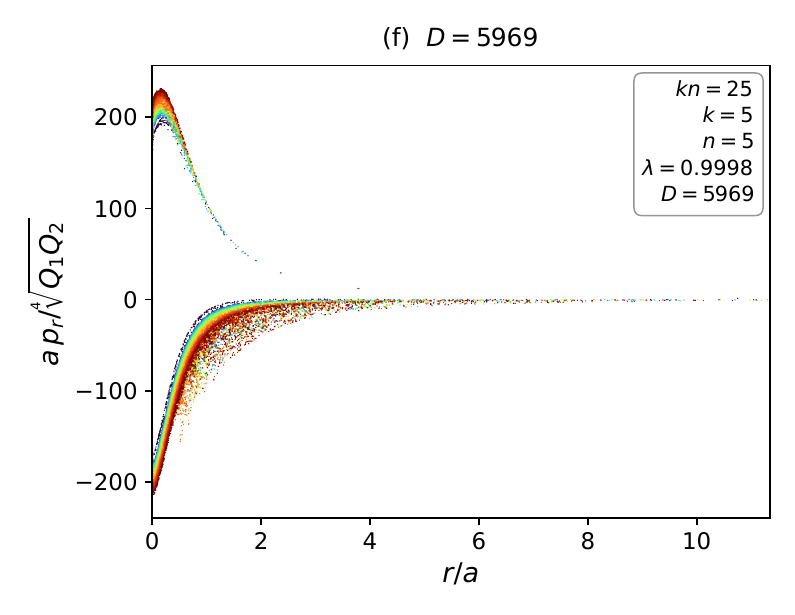}
\caption{Null-geodesic Poincar\'e sections of the cohomogeneity-3 $\left(Q_1 \neq Q_2\right)$ momentum-fractionated Style-1 superstratum as the throat is deepened. The surface of section is $\theta = \pi/4$ with $p_\theta >0$ on the null shell $H=0$, projected onto $\left(r/a, a p_r / \sqrt[4]{Q_1Q_2}\right)$. 
The geometry is now in the decoupling limit. The only quantity varied is $\lambda \equiv 2 b_4 /\sqrt{Q_1 Q_2}$, which fixes the cap scale and hence the throat depth $D \equiv \sqrt{Q_1Q_2}/a$: (a) $\lambda = 0.9,~D=274$; (b) $\lambda=0.99,~D=846$; (c) $\lambda =0.995,~D=1195$; (d) $\lambda =0.998,~D=1888$; (e) $\lambda = 0.99944,~D=3567$; (f) $\lambda =0.9998, ~D=5969$. Deepening the throat therefore fails to restore integrability at any finite $D$ accessible here, in contrast to the $Q_1=Q_2$ case, whose sections consist of nested invariant curves in this limit. All panels share the same charges and background data: $Q_1 = 54.66$, $Q_2 = 16.46$, $R = 8.441$, $k=n=5$, $m_1=6$, $m_2=-4$, $p_u=p_v = 0.10994$.}
\label{fig:momntfracsscoh3}
\end{figure}

Poincar\' e sections are shown in fig.~\ref{fig:momntfracsscoh3}, where we always have a chaotic sea, in accordance with nonseparability, but this time the separability is not restored even in the infinite throat limit, which we see in fig.~\ref{fig:momntfracsscoh3} as the existence of a narrow chaotic layer even for very large $D$. The reason is that the two families have different obstructions to integrability. For the $Q_1=Q_2$ case, the obstruction is the nonseparability of the effective potential in $(r,\theta)$, and it is suppressed as $a^2$. Splitting the charges adds a second, independent obstruction -- the $\zeta$-dependence carried by the connection $\omega_{r,\theta} \propto \mathcal C_1 = i b_4 \left( \chi - \chi^{-1} \right)$ (see eq.~(\ref{eq:qgenomegar})). Its amplitude is $|\mathcal C_1| \to \lambda \sqrt{Q_1 Q_2} \sinh \bar s$, where $e^{2\bar s}\equiv Q_1/Q_2$ parametrizes the charge asymmetry. This amplitude is controlled by $\lambda$ and $\bar s$ and not by $a$, and $\zeta$-dependence will thus persist as throat gets deeper. That being said, in the strict $a \to 0~(D \to \infty)$ limit we also have to take into account the fact that $\omega_{r,\theta} \propto \Delta^{kn}$, so $r$ and $\theta$ legs of $\omega$ vanish as $a \to 0$, and with them we lose $\zeta$-dependence too.

Now that we have classified the separable and nonseparable cases of both $Q_1=Q_2$ and $Q_1\neq Q_2$ geometries, we see that separability is an exception; the generic situation is nonseparable. The corresponding Poincar\' e sections thus generically show geodesic chaos, but they also show that chaos eventually weakens and eventually disappears when the throat $\sqrt{Q_1Q_2}/a$ becomes arbitrarily long (similar to the findings of \cite{Djukic:2026ejk}). The crucial issue is how this squares with the wave chaos, that directly connects to the structure of holographic states (in the extended dictionary picture).

\section{Extending the Random Wave Chaos Hierarchy}\label{secwaves}

\subsection{Berry Random Wave Conjecture}
\label{sec:BRWtests}

We start the section on wave chaos with a short summary on the Berry random wave hypothesis and how exactly it relates to geodesic motion and separability. Although all of this is known from the literature, we feel that a few comments are in order, to make the context of our findings clear, as there is some confusion in the literature about what the Berry random wave criterion really tests.

The Berry conjecture \cite{Berry:1977} states that an eigenfunction supported on a classically chaotic region locally looks like a superposition of plane waves with fixed momentum $k$ and random phases. One can easily compute the 2-point correlation functions of such a random superposition of plane waves -- the general expression was reviewed in the Introduction (eq.~(\ref{eq:brw})). Specializing to systems with $\mathfrak D=2$ degrees of freedom, the correlator (\ref{eq:brw}) becomes:
\be
C_2(\vec x_1, \vec x_2)\equiv C_2(\vec x_1-\vec x_2) = J_0(k |\vec x_1 - \vec x_2|).\label{eq:berry2}
\ee
The original argument by Berry concerns the Wigner pseudodistribution function
\be
    W(\vec p, \vec x) = \frac{1}{\left(2\pi \hbar \right)^2} \int d^2 q ~\psi \left( \vec x + \frac{\vec q}{2} \right) \psi^\dagger \left( \vec x - \frac{\vec q}{2} \right)~e^{-i \vec p \cdot \vec q/\hbar}
\ee
for some wavefunction $\psi$. The one assumption that enters Berry's derivation is what the Wigner function looks like semiclassically. For classically irregular/chaotic systems the Wigner pseudodistribution equals the standard microcanonical measure of the energy shell: $W \sim \delta\left( H(\vec p, \vec x) - E \right)$. This is intuitively logical as we expect a strongly chaotic system to uniformly explore the whole energy hypersurface and to reach thermodynamic equilibrium. A more rigorous derivation was provided by Leboeuf and Voros \cite{Leboeuf:1990}. Computing the Fourier transform of the microcanonical $W$ as $C_2 (\vec x; \vec y) \sim \int d^2 p ~e^{i \vec p \cdot \vec y/ \hbar}~W(\vec p, \vec x)$ we get
\be
    C_2(\vec x; \vec y) = J_0 \left( k (\vec y) |\vec x| \right),\label{eq:berry1}
\ee
with $k(\vec y)= \sqrt{2m \left( E- V(\vec y) \right)}/\hbar$ for Hamiltonians of form $H = p^2/2m - V(\vec x)$. This is the same form as (\ref{eq:berry2}) except that now $k$ need not be constant in space anymore.

We should contrast this with the regular case, in which $W\sim \delta^{(2)}\left( \vec I(\vec p, \vec x) - \vec I_0 \right)$ -- the Wigner pseudodistribution is supported on classical invariant tori, and it is only nonzero where the actions (integrals of motion) $\vec{I}$, that can be found from the canonical coordinates and momenta $\vec{x}$ and $\vec{p}$, equal their values on the torus $\vec{I}_0$, yielding $C_2(\vec x; \vec y)=(1/2 \mathfrak D)\sum_{j=1}^{\mathfrak D}e^{i\vec p_j(\vec y) \cdot\vec x/\hbar}$, where for two degrees of freedom we have $\mathfrak D=2$. This is a textbook result found, e.g., in \cite{Wimberger:2022}. Therefore, in quantum systems whose classical limit yields a chaotic sea in the phase space, the momenta are uniformly distributed on $\mathbb{S}^1$, as opposed to having support on only $2 \mathfrak D$ points. This motivates the Berry conjecture, which predicts that the statistical structure of a chaotic wavefunction is captured by the uniform random sum:
\be
    \psi \left(\vec x \right) = \sum_{j=1}^N a_j \cos \left( k \hat n_j \cdot \vec x + \phi_j \right),\label{eq:berrysum}
\ee
where $\hat n_j$ are equidistant and $\phi_j$ are random uniform on $\mathbb{S}^1$. This directly implies the expression (\ref{eq:berry2}) upon averaging over the circle. The point is that the Berry conjecture tests the chaoticity of the semiclassical limit in phase space. It does not presume the existence of a Random Matrix Theory (RMT) description for the spectrum; such a spectrum is sufficient but not necessary to satisfy the random wave hypothesis. Therefore, our tests do not directly probe the RMT statistics of the Hamiltonian; they probe the existence of a semiclassically chaotic limit.

Apart from the correlation function, the other easy test of the Berry random wave behavior is the distribution function of the intensities. The sum of $N \gg 1$ independent, comparably-sized terms in (\ref{eq:berrysum}) via the Central Limit Theorem gives  Gaussian distribution for $\psi$. The change of variables from $\psi$ to the intensity $I =|\psi|^2$ leads to the Porter-Thomas law for the probability distribution of intensities $P_\mathrm{TP}(I)$. For systems with trivial time-reversal invariance it reads
\be \label{eq:PTlaw}
P_{\mathrm{TP}}(I)=\frac{1}{\sqrt{2\pi I}}e^{-\frac{I}{2}}.
\ee
In addition to plotting the histogram of the intensity and comparing to the Porter-Thomas prediction (\ref{eq:PTlaw}), another practical criterion to check how close our data are to the Porter-Thomas distribution is to compare the cumulative distribution function $w_<(I_0)$, i.e. the total weight up to some cutoff $I_0$, to the Porter-Thomas prediction, which is easily found as:
\be
w_<(I_0)=\int_0^{I_0}P(I)dI = \mathrm{Erf}\left(\sqrt{\frac{I_0}{2}}\right).\label{eq:PTlawcumul}
\ee
Both criteria (the correlation function (\ref{eq:berry2}) and the Porter-Thomas distribution (\ref{eq:PTlaw})) were used in \cite{Djukic:2026ejk} as diagnostics of wave chaos for BPS microstate geometries. The correlator probes the second moment and the isotropy of the wavefunction distribution; the intensity distribution probes Gaussianity. They are independent tests, and a mode can pass one and fail the other \cite{Djukic:2026ejk}.\footnote{For other tests of wave chaos in 2 dimensions, see \cite{Ageev:2025yiq,Ageev:2025iiy,Ageev:2026jvx}.}

A final remark is in order. While the Berry random wave hypothesis is about studying the \emph{wavefunctions} of a \emph{quantum-mechanical} system, here we actually study a massless scalar \emph{field} in a \emph{classical theory}. But this does not really present an obstacle. From the bulk (field) viewpoint, our classical wavefunction $\Phi$ is really the classical limit of the quantum field $\hat{\Phi}$ in the bulk, i.e., the eikonal limit of some combination of single-particle eigenstates of \emph{fixed energy}.\footnote{While we find it convenient to use the coordinates $u$ and $v$ which both mix time with a spatial direction, we could easily express the energy $\omega$ in terms of $p_u$, $p_v$.} Schematically, we can write $\Phi=\sum_na_n\vert u_n\rangle+\mathrm{h.c.}$, so the random-wave chaos really lies (mostly) in the high-$n$ modes $\vert u_n\rangle$. Since they all carry the same energy (or, alternatively, could be thought of as coming from a microcanonical ensemble with energies between $E-\Delta E$ and $E+\Delta E$, with $\Delta E\ll E$), the Berry hypothesis still holds. Here again it becomes clear that our procedure is similar in spirit but definitely inequivalent to the implementation of the LMRS logic from \cite{Chen:2024oqv,Chen:2026vml}, where one mixes the states with, in general, different energies.

\subsection{Wave chaos in cohomogeneity-2 solutions ($Q_1=Q_2$)}

Now we will apply the tests delineated above to our momentum-fractionated solutions. Let us start from the cohomogeneity-2 solutions given in (\ref{eq:metsty2}). For this metric, the symmetry-consistent ansatz for a scalar wave is:
\be
\Phi=e^{i\left(p_uu+p_vv+m_1\varphi_1+m_2\varphi_2\right)}f(r,\theta).
\ee
The massless Klein-Gordon equation $\Box \Phi =0$ now takes the form
\be
\frac{1}{r}\partial_r\left[r\left(r^2+a^2\right)\partial_rf\right]+\frac{1}{\cos\theta\sin\theta}\partial_\theta\left(\sin\theta\cos\theta\partial_\theta f\right)+\Sigma \mathcal W f = 0,\label{eq:kg}
\ee
where we have again
\be
    \mathcal W = \mathcal P \left(2  p_u p_v - \mathcal F p_u^2 \right) - \frac{M_1^2}{(r^2 +a^2)\sin^2 \theta} - \frac{M_2^2}{r^2 \cos^2 \theta}, \quad M_i = m_i - \beta_i p_v - \omega_i p_u.\label{eq:ww}
\ee
Since $\sqrt{\mathcal P}$ cancels out of the derivative terms completely, the derivative part is already split into an $r$-only and a $\theta$-only operator with no cross terms and no residual conformal factor. So, the massless wave equation separates if and only if $\partial_r \partial_\theta(\Sigma \mathcal W)=0$, which is precisely the condition for the null-geodesic Hamilton-Jacobi separability. There is a general lesson from quantum chaos that is nicely stated in \cite{Haake:book} (p.~82): \emph{``The common origin of quantum and wave chaos may be seen in the nonseparability of the wave equation, just as we may attribute chaos in classical mechanics to the nonseparability of the Hamilton-Jacobi equation.''}

We can further massage the wave equation into the Helmholtz form, suitable for performing the Berry random wave test, by introducing $r = a \sinh \xi$ and $\psi = f \sqrt{\sinh 2\xi \sin 2 \theta}$:
\be
\partial_\xi^2 \psi + \partial_\theta^2 \psi + k_{\mathrm{eff}}^2(\xi,\theta) \psi = 0, \quad k_{\mathrm{eff}}^2 = \Sigma \mathcal W+ \frac{1}{\sinh^2 2\xi}+ \frac{1}{\sin^2 2 \theta}.\label{eq:helmholtz}
\ee
Since $(\xi,\theta)$ is a flat 2-dimensional coordinate patch, the comparison to the Berry correlation function $C_{\mathrm{irreg}} = J_0 (k |\Delta|)$ $(\mathfrak D=2)$ can be made without any further ambiguity about what the ``distance'' $|\Delta|$ means, i.e. $\psi$ -- not $f$ -- is the field to which Berry's Gaussian-random-field statement rigorously applies.\footnote{In the literature, the Berry random wave criterion is still often applied on non-flat manifolds or for non-constant potentials \cite{Toscano:2002,Mirlin:2002,Urbina:2003xip}, in a form which is slightly generalized though less rigorous. While such approaches often work surprisingly well, it is of course simpler to perform the test on the billiard type problem (\ref{eq:helmholtz}) when we have the opportunity.} The numerical procedure for solving (\ref{eq:helmholtz}) is given in Appendix \ref{secappbb}. All statistics are restricted to the classically allowed region $k_{\mathrm{eff}}^2 > 0$. To check quantitatively the consistency of data with the Berry random wave paradigm, we have used the Kolmogorov-Smirnov test. We have also performed the rank-1 test of the eigenmodes to quantify how far from separability the equation is. Both tests are explained in more detail in Appendix \ref{secappc}.

\subsubsection{Boundary conditions}

In order to perform numerical calculations we also need the behavior on the boundaries, to feed the boundary conditions to the integrator. The reduced problem lives on $(r, \theta) \in (0, \infty) \times (0,\pi/2)$, so there are four edges. Three of them are identical in the asymptotically flat and decoupling description limits; only the UV differs, which corresponds precisely to the flat vs. AdS asymptotics.

The cap at $r \to 0$ gives $\Sigma \to a^2 \cos^2 \theta$, $\omega_2 \to 2 k R$ and $\beta_2 \to - 2 kR$ independently of $b_4$ and of whether the ``1'' in $\mathcal P$ is kept. The $M_2^2/r^2 \cos^2\theta$ term then dominates against $a^2(\partial_r^2 + r^{-1}\partial_r)$:
\be
\Sigma\mathcal W \to - \frac{a^2\left(M_2^{(0)}\right)^2}{r^2}, \quad M_2^{(0)}=m_2+2kRq, \quad q = p_v-p_u,
\ee
giving $f\sim r^{\pm\nu}$, $\nu=\left|M_2^{(0)}\right|$. Regularity selects the $r^\nu$ branch. This is the behavior at the (smooth) cap, so it could not have depended on the UV asymptotics.

On the other hand, $\theta \to 0$ gives $\omega_1, \beta_1 \propto \sin^2 \theta$, which vanish there. Likewise, for $\theta\to\pi/2$ we get $\omega_2,\beta_2 \propto \cos^2 \theta$, again vanishing. This implies $M_1 \to m_1,~M_2 \to m_2$, resulting in $f(\theta\to 0)\sim\theta^{\pm |m_1|}$ and $f(\theta\to\pi/2)\sim(\pi/2-\theta)^{\pm |m_2|}$. Regularity requires the $+$ branches. This again does not depend on the decoupling limit.

Now we consider the UV. Let us start from the decoupling limit $\mathcal P \to Q^2/\Sigma^2$ when at large $r$
\be
    \Sigma \mathcal W \to \frac{2 Q^2 p_u p_v}{r^2} - \left ( \frac{m_1^2}{\sin^2\theta} + \frac{m_2^2}{\cos^2 \theta} \right)\to \mathrm{const.},
\ee
the radial equation becomes $r^2R''+3rR'-l(l+2)R=0$ with $l$ being the angular momentum of the $\mathbb{S}^3$ harmonic, so
\be
    f \sim A_+ r^l + A_- r^{-(l+2)}.
\ee
These are the source and VEV branches of an AdS$\mathrm{}_3$ field of dimension $\Delta_f= l+2$. The boundary condition is $A_+ = 0$ (normalizability/no source). Due to the confinement and energy gap in global AdS, the spectrum is discrete and unbounded, meaning we have a genuine self-adjoint eigenvalue problem. The $b_4$-dependent piece behaves as $\mathcal O\left(r^{-4}\right)$ and does not influence the exponents.

On the other hand, in the asymptotically flat case, we have $\mathcal P \to 1$, $\mathcal F,\omega,\beta \to 0$, and
\be
    \Sigma \mathcal W \to 2 p_up_v r^2 - \left( \frac{m_1^2}{\sin^2 \theta}+ \frac{m_2^2}{\cos^2 \theta} \right).
\ee
The radial equation is now the Bessel equation $r^2 R''+3r R'+(\kappa r^2-l(l+2))R=0$, with $\kappa^2 = 2 p_u p_v$. The equation is solved by $R = r^{-1} Z_{l+1}(\kappa r)$. where $Z_l$ is a linear combination of Bessel functions defined as $Z_l (x) = c_1 J_l(x)+ c_2 Y_l(x)$, where $J$ and $Y$ are the usual Bessel functions of the first and second kind respectively, and $c_{1,2}$ are determined by boundary conditions. Now there are two possible boundary conditions:
\begin{itemize}
    \item $\kappa^2>0$, oscillatory at infinity -- continuum. There are no normalizable modes, only quasinormal resonances.
    \item $\kappa^2 <0$, which yields $R \sim r^{-1}e^{- |\kappa|r}$ -- a finite set of genuine bound states.
\end{itemize}

We argue that the decoupling AdS limit is natural for testing the wave chaos. Berry random wave hypothesis is a statement about bound eigenfunctions in a confined, classically chaotic region. In the asymptotically flat limit, this certainly excludes the $\kappa>0$ case, whose unbounded solution cannot show true chaos (at best it can show transient chaos). For $\kappa <0$, however, we only have a finite set of modes, all of them likely trapped deep in the throat. Such a situation, where the wave is confined to a small region where only a few low-lying modes can fit, will give poor statistics and will not probe most of the flat region. We conclude that the full metric with flat space asymptotics does not really probe chaos in a meaningful way. We therefore work in the decoupling limit. This is also the reason why in our previous work \cite{Djukic:2026ejk} we have taken the decoupling limit for all backgrounds.

\subsection{General case: cohomogeneity-3 solutions}

We take the following ansatz for the massless Klein-Gordon field:
\be
\Phi = \sum_m e^{i(p_u u + p_v v+m_1\varphi_1 +m_2\varphi_2)}e^{im \zeta}f_m(r,\theta),
\ee
where we have Fourier-expanded over $\zeta$ (defined in eq.~(\ref{eq:zeta})), and the mode $m$ carries effective charges $p_v^{(m)} = p_v + m/2R,~m_2^{(m)} = m_2 - mk$. The reduced wave equation then becomes
\be \label{eq:masKGcoh3}
    \hat{\mathcal L}\left(p_v^{(m)}, m_2^{(m)}\right)f_m + \hat \delta_m f_m  + \mathcal C_1 \hat{\mathcal T}^+_m f_{m-n}+\bar{\mathcal C}_1 \hat{\mathcal T}^-_m f_{m+n} = 0.
\ee
We will first explain the notation in (\ref{eq:masKGcoh3}). In order to define the operators $\hat{\mathcal{T}}_m^\pm$, $\hat{\delta}_m$, $\mathcal{C}_1$, $\bar{\mathcal{C}}_1$ and $\hat{\mathcal L}$, let us first introduce the block derivative vector operator in the coordinate patch
\be
D_\mu^{(m)}=\left(ip_u,i\left(p_v+\frac{m}{2R}\right),\partial_r,\partial_\theta,im_1,i\left(m_2-mk\right)\right).
\ee
From the spacetime viewpoint, this is a (dual) vector of derivative operators, with spacetime index $\mu$. From the viewpoint of the space of $\zeta$-modes (functions) $f_m$, it is however a scalar, as it acts on sectors with fixed $m$. Now we write the inverse metric in terms of $\zeta$-dependent Fourier blocks:
\be
\sqrt{-g}g^{\mu\nu}(r,\theta,\zeta)=A_0^{\mu\nu}(r,\theta)+A_{+n}^{\mu\nu}(r,\theta)e^{-in \zeta}+A_{-n}^{\mu\nu}(r,\theta)e^{in\zeta},\quad A_{-n}\equiv\bar{A}_{+n},
\ee
where $A^{\mu\nu}_{0,\pm n}$ are spacetime matrices (two Greek indices), and also matrices in the mode space. Writing out the covariant box operator of the Klein-Gordon equation in terms of $D_\mu^{(m)}$ and $A^{\mu\nu}_{0,\pm n}$, we get
\be \label{eq:D}
D^{(m)}_\mu A_0^{\mu \nu}D^{(m)}_\nu f_m + D_\mu^{(m)}A_{+n}^{\mu\nu}D_{\nu}^{(m+n)}f_{m+n}+ D_{\mu}^{(m)}A_{-n}^{\mu\nu}D_{\nu}^{(m-n)}f_{m-n} = 0.
\ee
Finally, putting the pieces together we find exactly the equation (\ref{eq:masKGcoh3}) with
\bea
&&\hat{\mathcal L}\left(p_v^{(m)}, m_2^{(m)}\right) \equiv D_\mu^{(m)}A_0^{(0)\mu \nu}D_\nu^{(m)}, \quad A_0^{(0)} \equiv A_0|_{\mathcal C_1 = 0}, \\
&& \hat \delta_m \equiv D_\mu^{(m)} \left[ A_0 - A_0^{(0)} \right]^{\mu\nu} D_\nu^{(m)} = O \left( |\mathcal C_1|^2\right), \\
&& \mathcal C_1 \hat{\mathcal T}^+_m \equiv D_\mu^{(m)}A_{+n}^{\mu\nu}D_\nu^{m+n}, \\
&& \bar{\mathcal C}_1 \hat{\mathcal T}_m^- \equiv D_{\mu}^{(m)} A_{-n}^{\mu\nu}D_{\nu}^{m-n}.
\eea
Notice that eq.~(\ref{eq:masKGcoh3}) is tridiagonal in steps of $n$, because the metric has one harmonic. At $Q_1=Q_2$ it reduces to the potential $\Sigma \mathcal W$ that we have used throughout the cohomogeneity-2 computations. 

Finally, the asymptotic behavior on the boundaries is the same as for the cohomogeneity-2 case for each individual $m$. Specifying the boundary conditions thus reduces to what we have already discussed in the previous subsection.

\subsection{Berry random wave tests}

We now apply the tests of Section~\ref{sec:BRWtests} to eigenmodes of the reduced Helmholtz problem (\ref{eq:helmholtz}) in the decoupling limit. Three diagnostics are used, which test the 2-point correlations, the Porter-Thomas distribution of intensities, and separability, respectively. Here we just list the tests, and delegate the details to Appendix \ref{secappc}:
\begin{itemize}
    \item comparing the correlator $C(s)$ against $J_0(s)$, with $s$ the appropriately rescaled separation (defined in eq.~(\ref{eq:s}) and explained below it);
    \item computing the statistical proximity (measured by the Kolmogorov-Smirnov distance, defined in eq.~(\ref{eq:ksdistance})) of the intensity distribution to the Porter-Thomas law~(\ref{eq:PTlaw});
    \item estimating how close a mode is to a product $F(\xi)G(\theta)$, i.e., how close the equation is to being separable (by computing the rank-1 fraction $\mathcal R_1$, defined in eq.~(\ref{eq:r1})).
\end{itemize}

\begin{figure}[ht]
\centering
(a)\includegraphics[width=.29\linewidth]{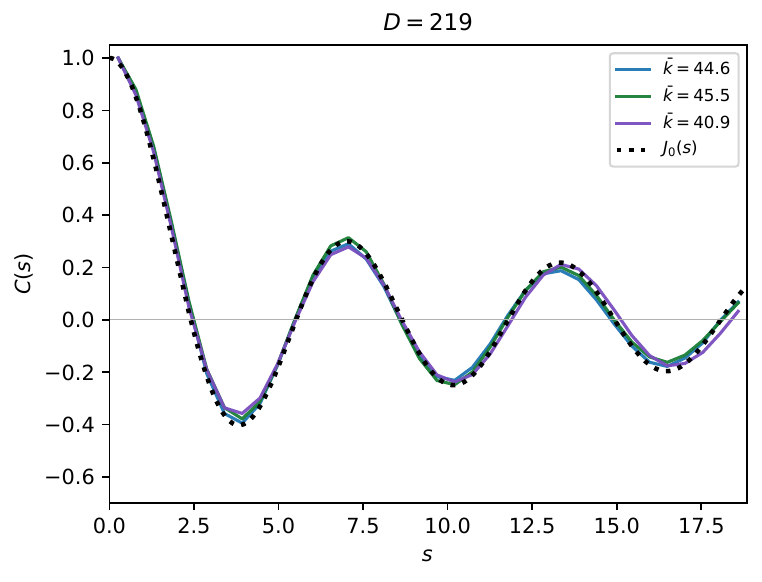}
(b)\includegraphics[width=.29\linewidth]{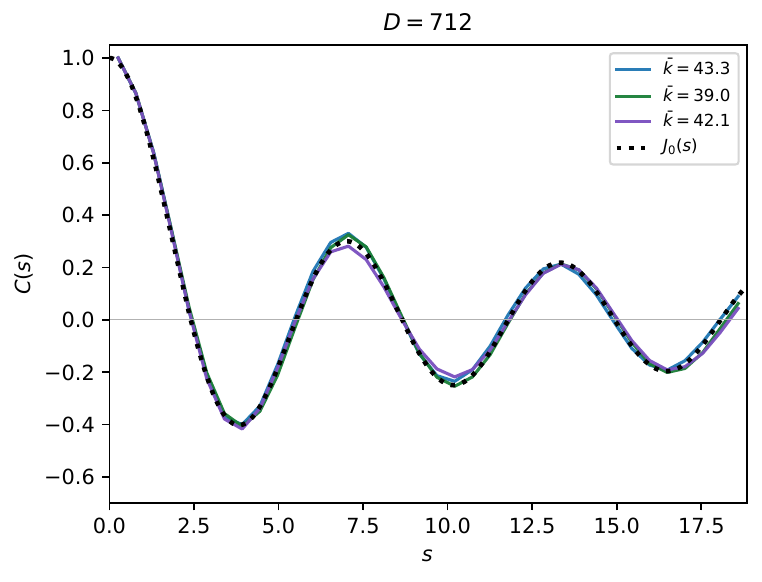}
(c)\includegraphics[width=.29\linewidth]{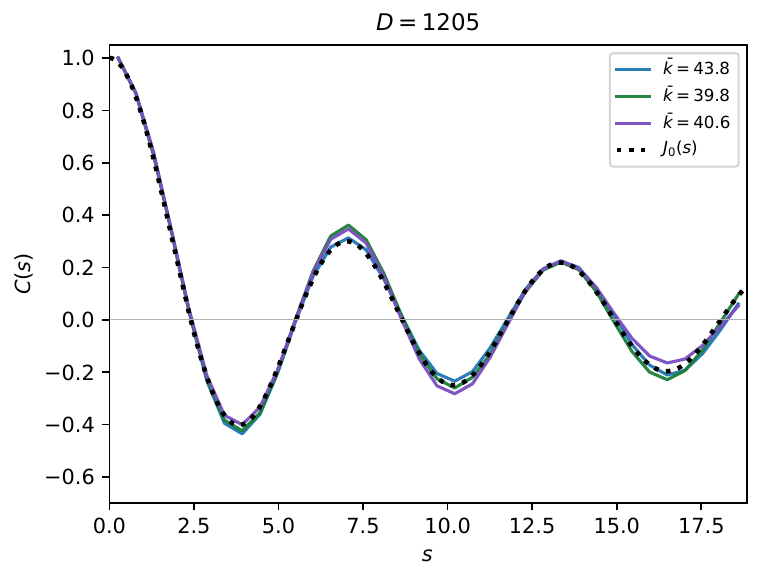}
(d)\includegraphics[width=.29\linewidth]{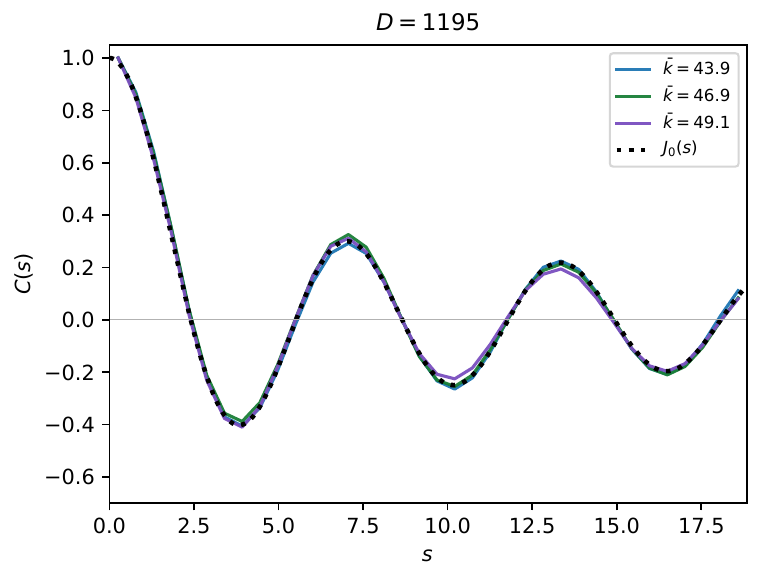}
(e)\includegraphics[width=.29\linewidth]{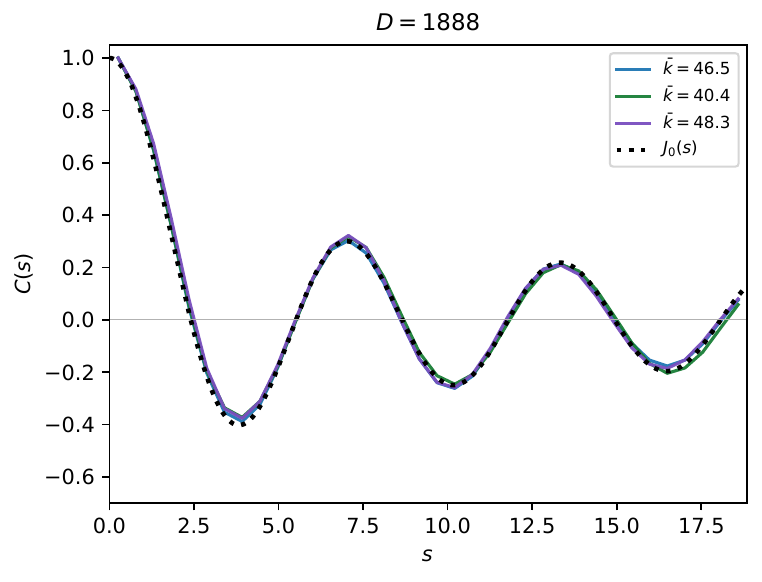}
(f)\includegraphics[width=.29\linewidth]{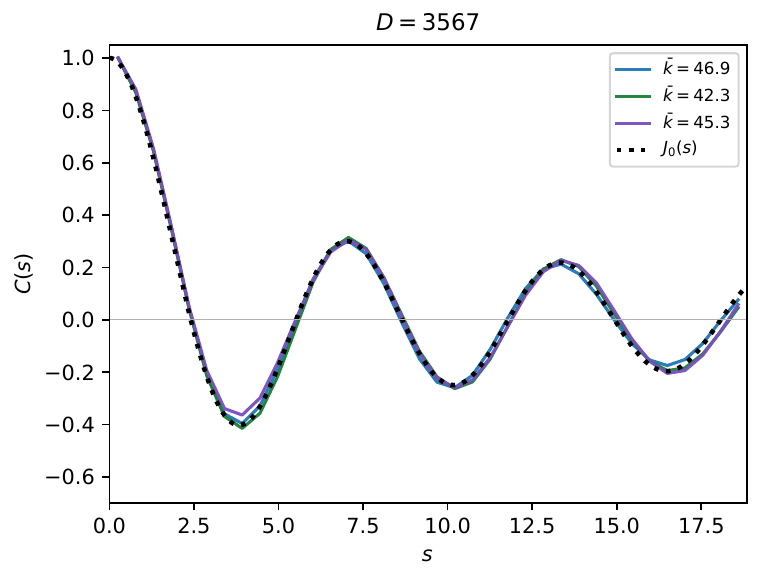}
\caption{The 2-point correlator of the eigenmodes of cohomogeneity-2 fractionated superstrata in the decoupling limit, $Q=30,~kR=8.441$, charges $(m_1,m_2)=(6,-4),~p_u=p_v$. Each panel shows the correlator $C(s)$ of the three most representative modes (thick colored) against the Berry prediction $J_0(s)$ (dotted); $s$ is the separation in units of local wavelength. Top row: approaching the $D\to\infty$ limit with $\lambda =0.9$ fixed and $(k,n) = (4,5),(13,5),(22,5)$, i.e., $kn = 20,65,110$ and $D= Q/a = 219$ (a), $712$ (b), $1205$ (c). Bottom row: approaching the $\lambda \to 1$ limit, with $n=k=5$ and $\lambda =0.995$ (d), $0.998$ (e), $0.99944$ (f) ($D = 1195,1888,3567$). The correlator follows $J_0$ at every throat length -- there is no sensitivity to the throat size.
}
\label{fig:momntfracsscorr1}
\end{figure}

\begin{figure}[ht]
\centering
(a)\includegraphics[width=.29\linewidth]{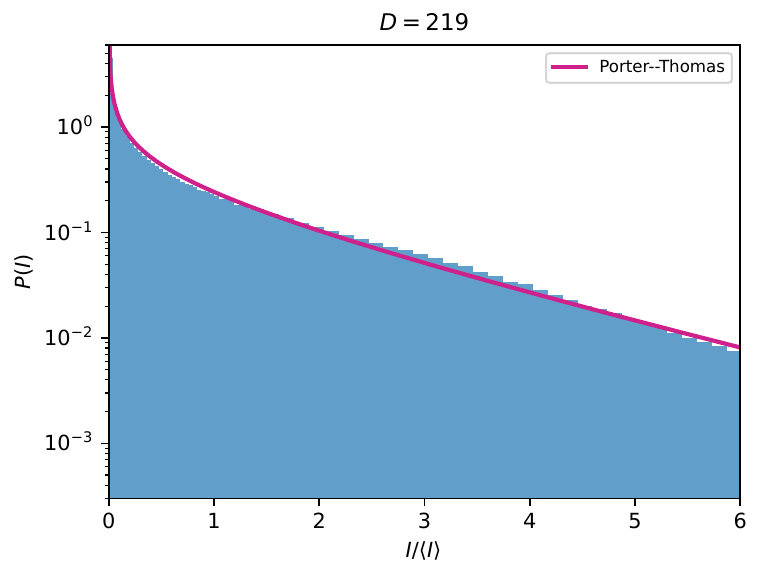}
(b)\includegraphics[width=.29\linewidth]{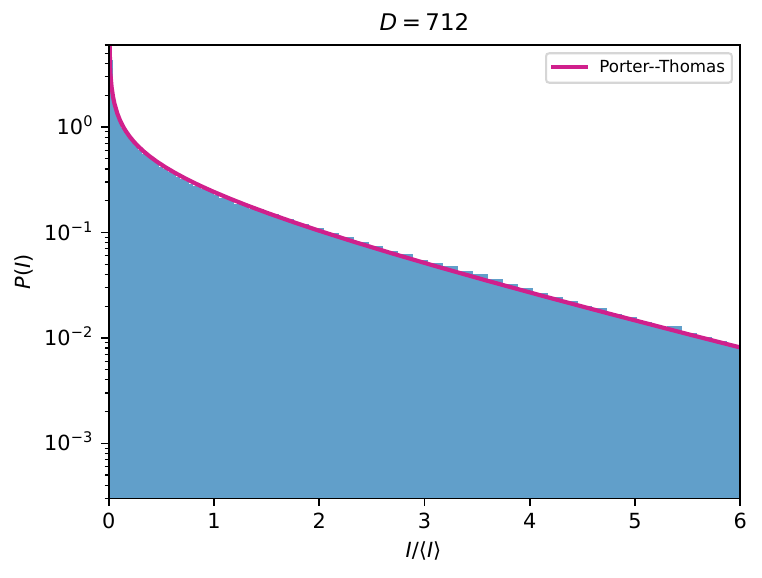}
(c)\includegraphics[width=.29\linewidth]{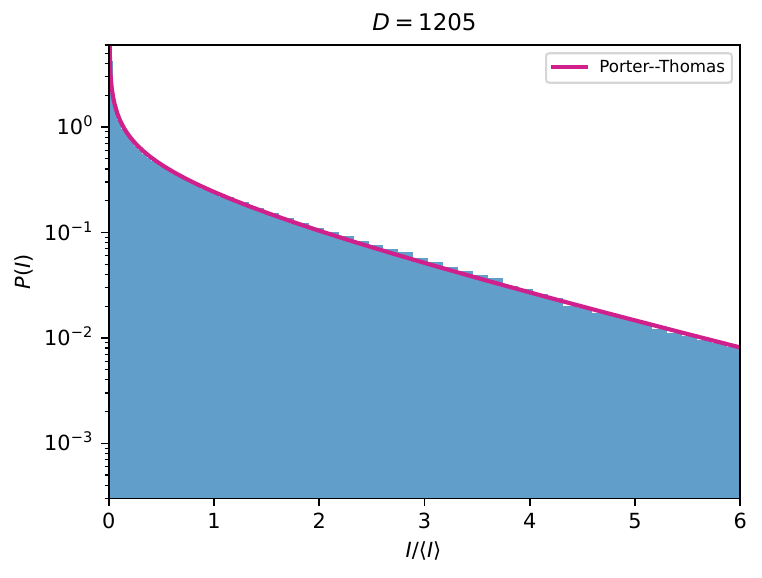}
(d)\includegraphics[width=.29\linewidth]{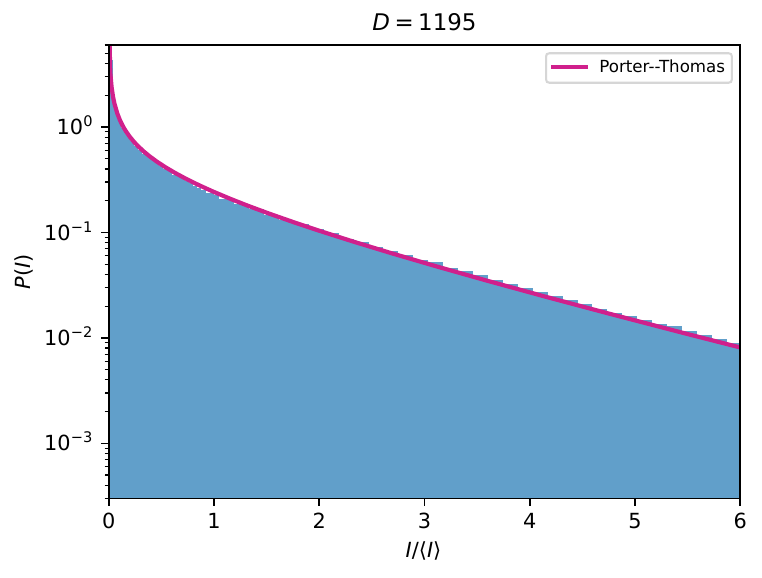}
(e)\includegraphics[width=.29\linewidth]{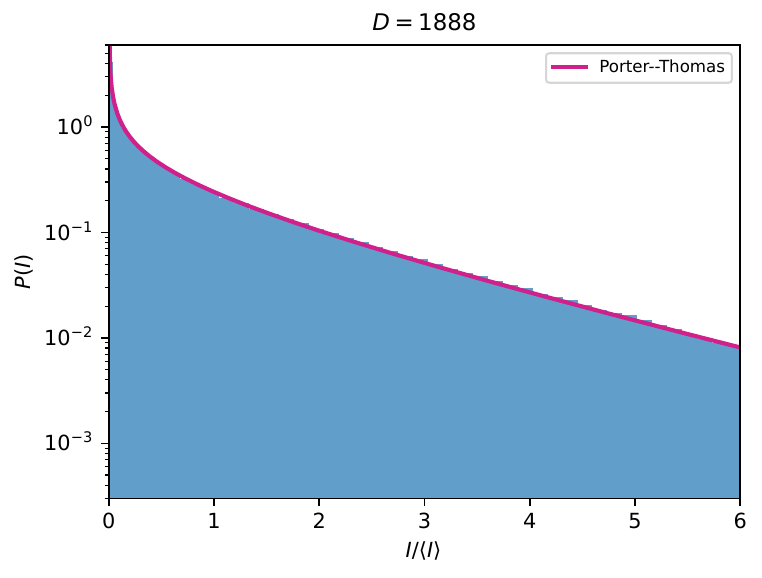}
(f)\includegraphics[width=.29\linewidth]{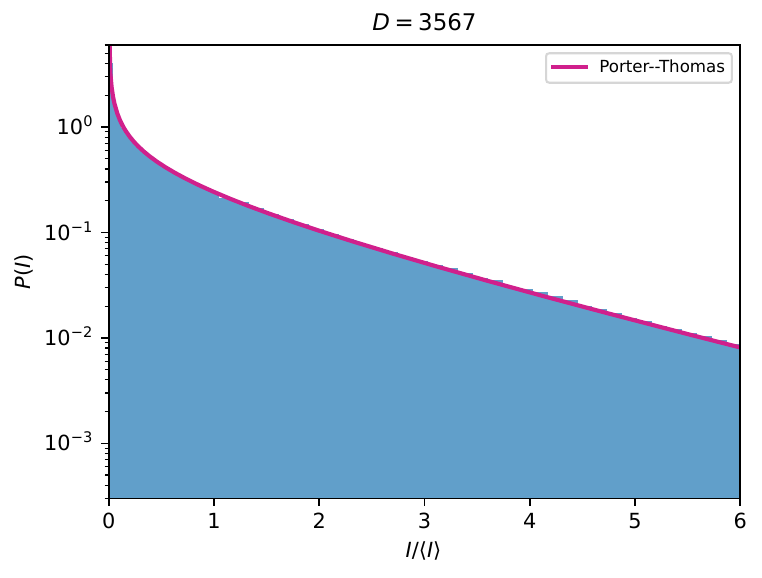}
\caption{Distributions of the normalized intensity $I/\langle I\rangle$ for the same backgrounds as in fig.~\ref{fig:momntfracsscorr1}, sampled over all $k$-modes, compared against the Porter-Thomas law (magenta), on logarithmic scale. Top row (a-c): the winding route at fixed $\lambda =0.9$ with $(k,n)=(4,5),(13,5),(22,5)$ and $D=219,712,1205$. Bottom row (d-f): $\lambda \to 1$ route at $(k,n)=(5,5)$ with $\lambda=0.995,0.998,0.99944$ and $D=1195,1888,3567$. Porter-Thomas holds in all cases, in agreement with fig.~\ref{fig:momntfracsscorr1}.
}
\label{fig:momntfracssPT1}
\end{figure}

Figs.~\ref{fig:momntfracsscorr1} and \ref{fig:momntfracssPT1} follow the cohomogeneity-2 backgrounds along the two routes of deepening the throat: increasing $k$ at fixed $\lambda$ (top rows) and approaching the scaling limit $\lambda \to 1$ at fixed $k$ (lower rows). The computed correlators follow the Berry form along both routes. The intensity histograms in fig.~\ref{fig:momntfracssPT1} show the same trend: panels (a-c) as well as (d-f) nicely follow Porter-Thomas law at every depth.

\begin{figure}[ht]
\centering
(a)\includegraphics[width=.29\linewidth]{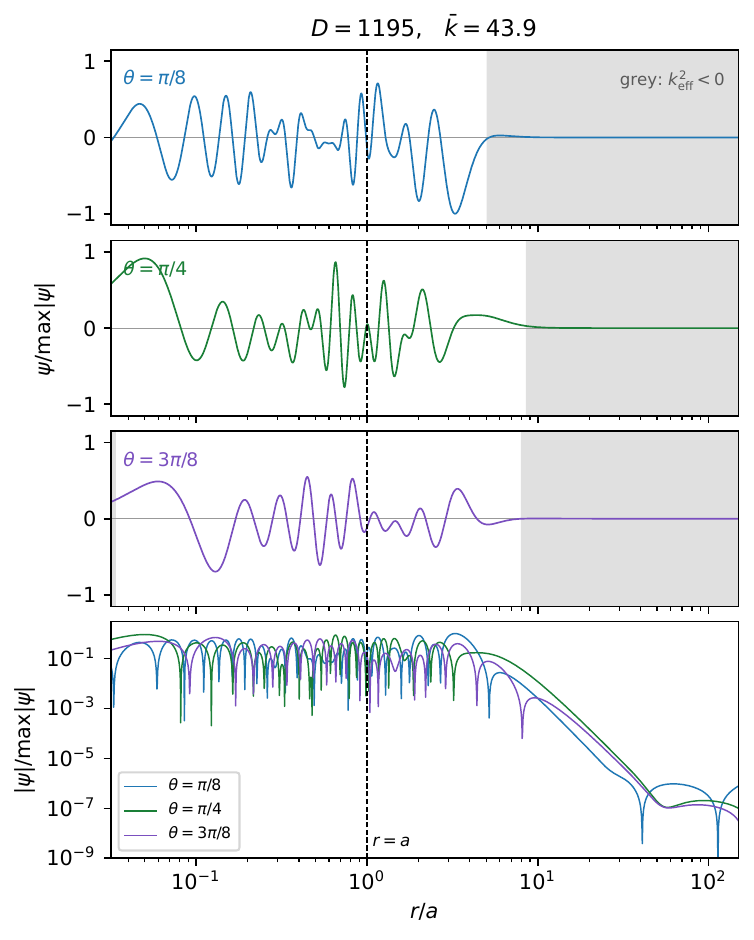}
(b)\includegraphics[width=.29\linewidth]{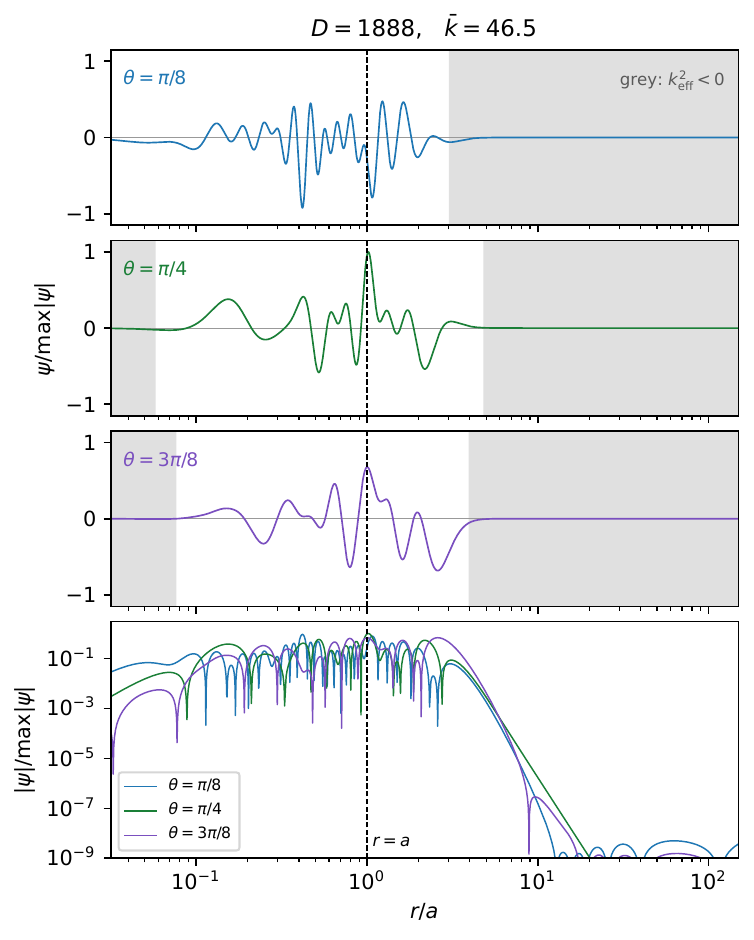}
(c)\includegraphics[width=.29\linewidth]{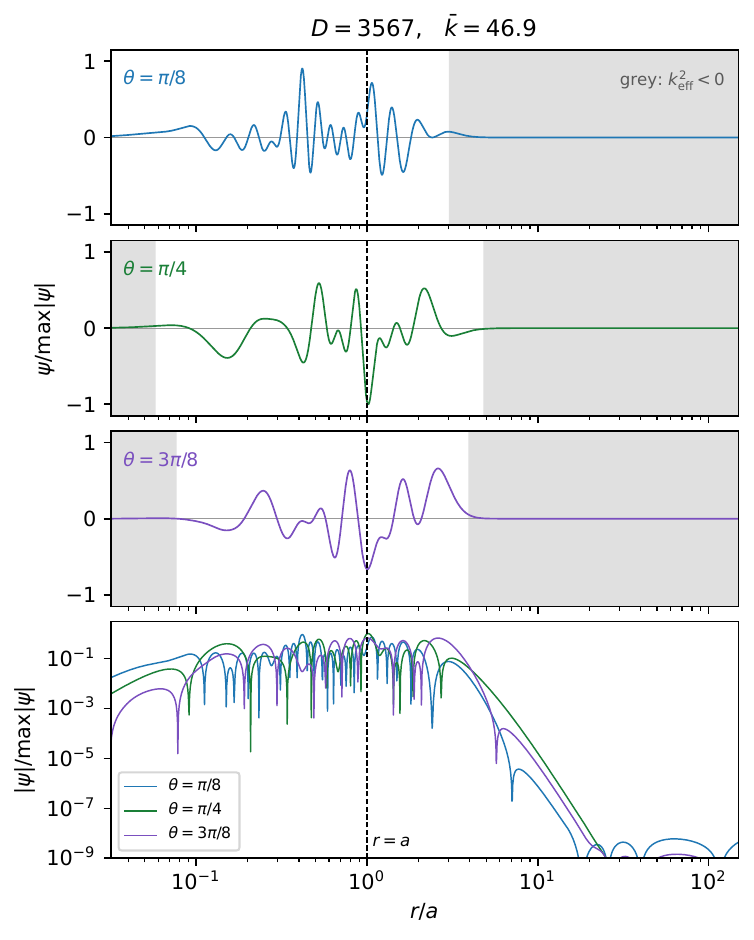}
\caption{Radial profiles of representative eigenmodes $\psi(r,\theta)$ in increasingly deep throats, for the same backgrounds as in fig.~\ref{fig:momntfracssPT1}(d-f). Each figure shows one $\bar{k}$-mode as a function of $\log r$, at three different polar angles $\theta$ (first three rows). The last row superimposes the previous three on logarithmic scale for $\psi$, with vertical dotted lines denoting the locations of their turning points. Gray bands mark the forbidden region $k_{\mathrm{eff}}^2<0$. The field oscillates throughout the cap and decays as soon as it crosses the turning point. Therefore, the wavefunctions almost do not explore throats longer than roughly $10a$.
}
\label{fig:momntfracradial}
\end{figure}

From the above, we conclude that the dynamics is insensitive to the depth of the AdS${}_2$ throat, at least for long enough throats. But increasing the throat depth indefinitely (along either route) yields the black hole limit (as the limit $D\to\infty$ or $\lambda\to 1$). Hence, one would expect that something important should happen in this limit, rather than nothing. The explanation is that the probe wave chaos analysis just cannot see the black hole limit properly. We will discuss the broader reasons for this in Section \ref{secconc}. Now we just give a formal argument on why the throat length does not matter, as long as it is above some threshold. The insensitivity of chaos to $D$ is expected from the general behavior of the radial profile of eigenfunctions in superstrata backgrounds \cite{Bena:2018bbd}: they oscillate in the cap and decay as soon as they enter the AdS${}_2$ region (see fig.~\ref{fig:momntfracradial}), as expected from the well known fact that AdS${}_2$ does not support finite-energy excitations \cite{Maldacena:1998uz}.\footnote{We thank Iosif Bena for pointing out the role of AdS${}_2$ region in confining the modes to the cap.} The modes therefore live almost entirely in the cap, and the cap is also where the nonseparable part of the potential sits: the obstruction $\frac12 \mathcal C \mathcal F$ in eq. (\ref{eq:potfinal}) is of order one only for $r\lesssim a$. The support of the waves and the source of their chaos are thus both confined to the cap region. There is some mixing between the cap and the throat, and for relatively small $D$ this mixing will influence chaos. But deepening the throat further just enlarges the part of space that modes never reach, so the random wave statistics \emph{cannot} depend on $D$. This also explains why the geodesics shown in figs.~\ref{fig:momntfracssPS}-\ref{fig:momntfracsscoh3} \emph{do} depend on $D$: unlike waves, which are delocalized, typical geodesics will localize (i.e., spend a long time) near the regular periodic orbits in the throat.

\begin{figure}[ht]
\centering
(a)\includegraphics[width=.45\linewidth]{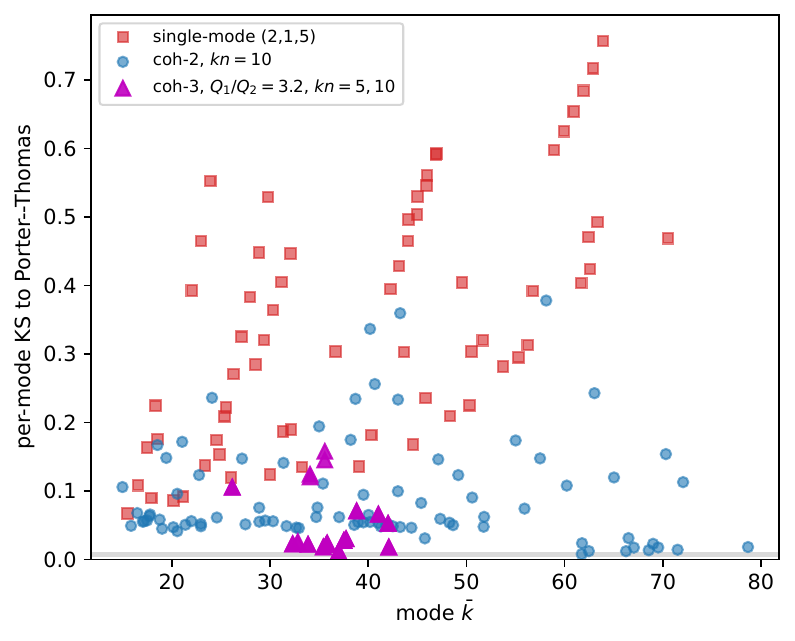}
(b)\includegraphics[width=.45\linewidth]{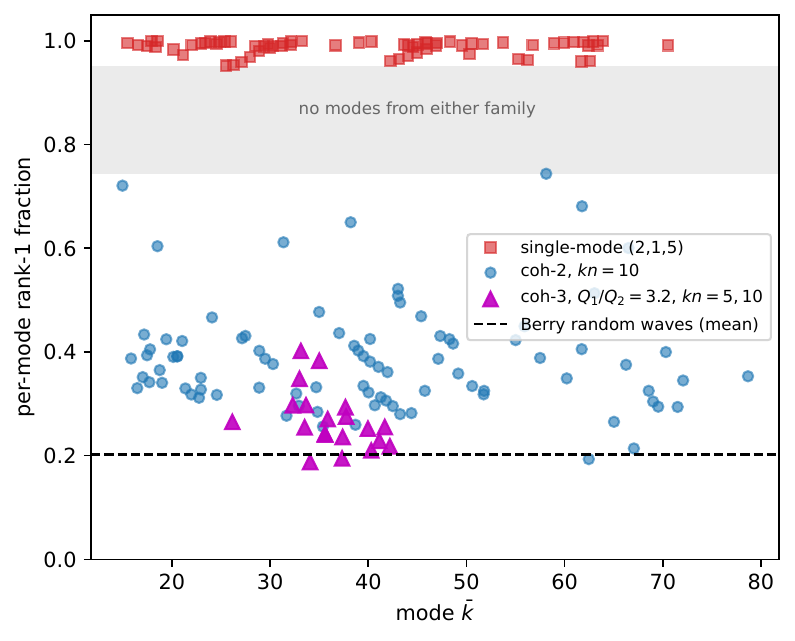}
\caption{Agreement with the Porter-Thomas intensity distribution (a) and the degree of nonseparability (b) for several $\bar{k}$-waves from each of the three families studied. Red squares stand for single-mode $(2,1,5)$ superstrata at depth $D=\sqrt{Q_1Q_5}/a = 25$; blue circles stand for cohomogeneity-2 fractionated superstrata at $kn=10, \lambda = 0.9, D=Q/a = 55$; magenta triangles are for cohomogeneity-3 fractionated superstrata with $Q_1/Q_2=3.2$ at $kn=5$ and $kn=10$ with $\lambda=0.9$ ($D=55$). In (a) we plot the Kolmogorov-Smirnov distance of each wave's intensity distribution from the Porter-Thomas law. The gray band is the reference value for an ensemble of random Gaussian fields. In (b) we give the rank-1 fraction for each $\bar{k}$, which is 1 for an exact product $f(\xi)g(\theta)$ and $\sim 0.2$ for an ensemble of Gaussian random waves (dashed line). The single-mode solutions are never far from separability, clearly different from the probes on fractionated backgrounds. The shaded band contains no mode of either family: the two families are well separated.}
\label{fig:momntfracKSrank_1}
\end{figure}

The weakening of geodesic chaos and the initial growth followed by saturation of the wave chaos with increasing $D$ were also found for non-fractionated geometries in \cite{Djukic:2026ejk}. In that work, we have focused on smaller values of $D$ where the initial growth of wave chaos with $D$ was still noticeable, although, e.g., in figs.~4 and 8 one can already see the saturation. But the level of saturation ($\chi^{-2}$, which is related, though not identical, to the Kolmogorov-Smirnov tests performed here) is below the maximum chaos level reached for momentum-fractionated backgrounds, as we show in fig.~\ref{fig:momntfracKSrank_1}. Therefore, the single-mode results of \cite{Djukic:2026ejk} harmonize with the fractionated results here.

The comparison with the waves on single-mode $(2,1,n)$ superstrata \cite{Bena:2016ypk} in fig.~\ref{fig:momntfracKSrank_1} places every mode of the cohomogeneity-2, cohomogeneity-3 and single-mode $(2,1,n)$ families on common axes, against the \emph{mean} wavenumber over the allowed region $\mathcal M$:
\be
    \bar k = \left( \frac{\int_{\mathcal M} k^2 \psi^2}{\int_{\mathcal M}\psi^2} \right)^{\frac12},
\ee
so that the comparison is made mode by mode and not through a few chosen examples. In the Porter-Thomas test (fig.~\ref{fig:momntfracKSrank_1}(a)), the fractionated modes, whether cohomogeneity-2 or cohomogeneity-3, lie much closer to the reference level set by Gaussian random waves at every $\bar k$, while the $(2,1,n)$ modes lie an order of magnitude higher and drift further away as $\bar k$ increases, i.e. the semiclassical limit takes them away from the random wave behavior, rather than towards it. The separability test (rank-1 fraction $\mathcal R_1$) shown in fig.~\ref{fig:momntfracKSrank_1}(b) shows why: every single-mode eigenfunction, at all depths $D$ and all values of $\bar k$, is ``almost separable'': it can be approximated by a product of a radial and an angular function within a few percent, whereas the fractionated eigenfunctions have much lower rank-1 fraction, i.e., their nonseparability is much higher (though both cases are nonintegrable and chaotic, but to differing extents). The two families thus have clearly distinct dynamics, no matter what the throat depth or the wavenumber is.

The origin of the difference between non-fractionated and fractionated geometries lies in the structure of their potentials, namely, in how the nonseparable part of the effective
potential scales with the momenta. For the $(2,1,n)$ family, $\Sigma \mathcal W$ splits as $U_0 +\widetilde \lambda U_1+\widetilde \lambda^2 U_2$, where $\widetilde \lambda = p_u = p_v$. The $\widetilde \lambda^2$ term depends on $r$ only, $U_0$ is separable, and the entire obstruction to separability sits in the term $U_1$, linear in $\widetilde \lambda$. This is the obstruction found in \cite{Bena:2017upb}. Converting to our notation, the relative weight of the nonseparable term in $k_{\mathrm{eff}}^2$ is therefore $\sim (m_1+m_2)/ \widetilde \lambda \sim 1/\bar k$, so as $\bar k$ grows at fixed charges, the problem flows towards the integrable locus and the modes become closer and closer to product states. This is precisely what we see in fig.~\ref{fig:momntfracKSrank_1}(a). For the fractionated backgrounds, the obstruction is the term $\mathcal C = 8 k R p_u (m_1+m_2+2kR p_v)$ in eq.~(\ref{eq:potfinal}), i.e. the term quadratic in the momenta. At $p_u=p_v = \widetilde \lambda$, the term $\frac{1}{2} \mathcal C \mathcal F \simeq 8k^2 R^2 \widetilde{\lambda}^2 \mathcal F$ competes with the separable term $2\widetilde \lambda^2\mathcal P$ at every $\bar k$, with a $\widetilde \lambda$-independent relative weight set by $k^2 R^2 |\mathcal F|$, which is of order one in the cap. The eikonal limit thus never approaches the separable loci. The outcome is that single-mode superstrata become closer to integrable in the semiclassical limit, while fractionated ones do not.

\section{Fortuity of the dual CFT state?}\label{secfort}

The results of the previous section establish a sharp hierarchy: probe waves on single-mode superstrata remain close to separable and moderately chaotic, while on momentum-fractionated backgrounds they are far from separability and highly chaotic. It would be very appealing to trace this dichotomy back to the nature of the dual CFT states, as it is already known that single-mode superstrata correspond to monotone BPS states -- we will review this argument below. The more difficult part is to show that fractionated superstrata correspond to fortuitous states. Such a correspondence would turn the random-wave diagnostic into a supergravity-regime counterpart of the monotone/fortuitous classification \cite{Chang:2024zqi}, and would make concrete the expectation, outlined in the Introduction, that fractionation is the large-$N$ face of fortuity.\footnote{We are grateful to Rapha\" el Dulac for discussion on the fortuity of the dual CFT states, which has shaped the argument of this section.}

The dual CFT state of a single-mode superstratum at weak coupling and finite $N$ is given in \cite{Bena:2017xbt}, whereas the CFT state (in the same regime) for a fractionated superstratum is found in \cite{Bena:2016agb}. We first consider the states of single-mode superstrata. We will demonstrate a simple argument that they are monotone by increasing the rank $N=N_1N_5$ by a single unit. The state reads
\bea
&&\vert(k,m,n)\rangle=\left(|++\rangle_1\right)^{N_{++}} \left(\left(J_{-1}^+ \right)^m \left( L_{-1}-J_{-1}^3 \right)^n |00\rangle_{k} \right)^{N_{k,m,n}}\label{eq:windingconstrsinglemode}\\
&&N_{++}+ kN_{k,m,n} = N, \label{eq:windingconstrsinglemode1}
\eea
where the bottom line is the constraint to be satisfied by the occupation numbers. Now the strategy is to modify the state in such a way that $N$ (the total number on the right-hand side of (\ref{eq:windingconstrsinglemode1})) grows to exactly $N+1$. For the state (\ref{eq:windingconstrsinglemode}) this can only be achieved by adding a string of unit length, i.e., a single Ramond-Ramond vacuum strand $|++ \rangle_1$. In that case $N_{++} \mapsto N_{++}+1 \equiv \tilde N_{++}$, and $N\mapsto N+1$, and the winding constraint (\ref{eq:windingconstrsinglemode1}) then implies that $kN_{k,m,n}$ is unchanged. This is the only possible way to increase $N$ by one: if we fiddled with $N_{k,m,n}$ instead, the constraint (\ref{eq:windingconstrsinglemode1}) implies that $N$ would change by a larger amount. The structure of the resulting state with $\tilde{N}_{++}+1$ is the same as the original one, and since it is a BPS state we conclude that it has to be a monotone, graviton-gas type state, as explained in \cite{Bena:2025pcy}. 

Now consider the momentum-fractionated superstratum and its constraint \cite{Bena:2016agb}: 
\bea
    &&\left( |++ \rangle_k\right)^{n_1} \prod_{\hat m, \hat n, \hat p} \left( \left( J_{-1/k}^+ \right)^{k \hat p} | 00 \rangle_{k^2 \hat p} \right)^{n_{2,\hat p}} \left( \left( J_{-1/k}^+ \right)^{k \hat n} | ++ \rangle_{k(k \hat n+1)} \right)^{n_{3,\hat n}} \times \nonumber \\ 
    &&~~~~~~~~~~~~~~~~~~~~~~ \times\left( \left( J_{-1/k}^+ \right)^{k \hat m} | -- \rangle_{k(k \hat m-1)} \right)^{n_{4,\hat m}}\label{eq:fractstate}\\
    && k n_1 + k^2 \hat p n_{2,\hat p} + k (k \hat n+1) n_{3,\hat n}+ k(k \hat m -1) n_{4, \hat m} = N.\label{eq:fractstate1}
\eea
The background strands have length $k$ and the excited strands have lengths $\{k^2 \hat p,~k(k \hat n + 1),~k(k\hat m-1)\}$, which are fixed by a per-strand $\mathbb Z_\kappa$ condition, so appending a unit strand does not produce a member of the same family at $N+1$ but at best at $N+k$. Modifying other quantum numbers likewise increases $N$ by more than $1$. Therefore, monotonicity in the usual sense cannot be established by this construction. 

The arguments above are at the level of the free orbifold point strand structure. In order to make the above reasoning rigorous and determine whether the state is fortuitous (\ref{eq:fractstate}), one should consider the deformed $Q$-cohomology in the style of \cite{Chang:2025rqy}. This computation is beyond the scope of this paper.\footnote{Detailed computation is likely much more demanding than the sketchy argument that we provide here. In general, the fortuity properties need not be simply related to the structure of strands or cycles and how these change with $N$, as shown in \cite{Chae:2026vgn}.}

On the other hand, it was shown in \cite{Gaberdiel:2023dxt} that the worldsheet theory on $\left( \mathrm{AdS}_3 \times \mathbb{S}^3 \right) / \mathbb Z_k$ is the same free orbifold CFT expanded around a different background state, in which all $N$ copies are joined into $N/k$ $k$-wound strands in their ground state. This makes adding $|++\rangle_k$, instead of $|++\rangle_1$ as we did above, a more natural choice. The state (\ref{eq:fractstate}), regarded as an excitation of this theory, will now survive and is thus monotone in this generalized sense (we could say ``$k$-monotone'').\footnote{We thank Rapha\" el Dulac for pointing out the relevance of the $k$-wound vacuum to the notion of monotonicity used here.} But in this case $N$ jumps in units of $k$, so this is a modification of the very notion of monotonicity. Fractionated superstrata are thus ``$k$-monotone'' with respect to the $k$-wound RR vacuum above. In the light of the random wave chaos findings of this paper -- despite the fact that these states might or might not be fortuitous, and despite the known fact that they are still atypical microstates -- we regard them as useful toy models for black hole microstates that capture the physics of fractionally-moded microstates.

\section{Discussion and conclusions}\label{secconc}

We set out to ask whether momentum fractionation -- the microscopic mechanism which accounts for the entire gap between the entropy of a graviton gas and that of a black hole -- leaves a trace in the chaotic dynamics of probe waves on the corresponding supergravity backgrounds. The answer turns out to be positive -- fractionated microstate geometries systematically show much stronger chaos than non-fractionated ones. This goes hand in hand with the fact that 3-charge black holes (with finite horizon area) always have momentum fractionation, and agrees with the intuition (motivated in the Introduction) that fractionation is the large-$N$ analogue of fortuity.

The eigenmodes of the massless Klein-Gordon operator on the fractionated backgrounds, in the decoupling limit, are close to Gaussian random waves on both tests: the 2-point correlation function follows $J_0(s)$ and the Kolmogorov-Smirnov distance of the intensity distribution from the Porter-Thomas law is systematically much smaller than for the single-mode family. In our previous work \cite{Djukic:2026ejk} we found a trend of increasing wave chaos across families of decreasingly supersymmetric backgrounds whereas here we find a separation between two classes of the same family. This likely stems from the fact that the single-mode superstratum state is monotone in the sense of \cite{Chang:2024zqi}: adding one singly-wound RR vacuum strand takes it to a member of the same family at $N+1$. The fractionated superstratum, on the other hand, might be fortuitous -- there is no argument preventing it. A complete proof or disproof of fortuity requires detailed study of the deformed $Q$-cohomology \cite{Chang:2025rqy}, which we have not attempted.

Here we will comment on a puzzle we have encountered in connection to the black hole limit. The novelty of the fractional mode family of superstrata solutions as opposed to single-mode ones is that the throat can go deeper in two ways: either by increasing $k$ at fixed $\lambda$ (\ref{eq:lambdadef}), or by approaching the scaling limit $\lambda\to 1$ at fixed $k$. Along both routes, the random wave statistics are unchanged as $D$ grows: the modes oscillate in the cap and decay at the turning point.

The conundrum is that both routes lead towards the black hole: as the throat deepens the cap recedes, and in the infinite-$D$ limit one is left with the extremal BTZ$\times\mathbb{S}^3$ geometry whose near-horizon region is the AdS${}_2$ throat. A naive expectation would then be that wave chaos disappears whenever the black hole is approached (by either of the two routes in our case), in accordance with the emergent conformal symmetry of black hole throats. But now we will argue that our chaos probes cease to make sense in the black hole limit.

Namely, the probe approximation fails when $D\to\infty$ (along any route). A quantum of fixed energy at the AdS${}_3$ boundary arrives at the bottom of a deep throat with a blue shift that grows with $D$, and the tidal forces it experiences grow accordingly \cite{Tyukov:2017uig}. Eventually, the probe backreacts or excites stringy modes \cite{Bena:2020iyw}. For extremal horizons the problem is even sharper: generic small perturbations produce large or singular tidal forces at the horizon \cite{Horowitz:2023xyl}. Thus, a minimally coupled scalar or a null geodesic on a rigid long AdS${}_2$ throat is not a controlled approximation.\footnote{As a side remark, it is well known by now that the rigid AdS${}_2$ throat is itself not a consistent description of a black hole at lowest energy scales, because of the presence of the Schwarzian mode, which becomes strongly coupled below a scale set by the entropy \cite{Preskill:1991tb,Ghosh:2019rcj,Iliesiu:2020qvm,Iliesiu:2022onk,Kapec:2023ruw,Kapec:2024zdj}, and in the supersymmetric case this produces a gap above the exactly degenerate ground states \cite{Heydeman:2020hhw,Boruch:2022tno}.}

We conclude that \emph{``no probe chaos in the black hole limit''} is not a statement about black holes not being chaotic, but a statement that the probe wave/probe geodesic diagnostic becomes blind in the case of (near-)extremal black holes.

There is a more basic objection to everything we have done (here and in \cite{Djukic:2026ejk}), an obvious elephant in the room, which is not specifically related to (near-)extremal black holes, but just any black hole in general. The black holes themselves -- Kerr-Newman, the Myers-Perry and Kerr-NUT-(A)dS families in all dimensions, BTZ, and the non-extremal D1-D5-P black strings -- have completely integrable geodesic flow and a separable Klein-Gordon equation. This is due to the presence of the Carter constant \cite{Carter:1968rr} and its higher-dimensional generalization through the principal conformal Killing-Yano tensor \cite{Frolov:2017kze}. Applied to a black hole, every probe criterion we have used would return ``integrable''. In fact, the situation is worse: with ingoing boundary conditions at the horizon there is no discrete spectrum at all, only quasinormal modes, and the Berry conjecture, which is a statement about bound states, does not even apply. Yet black holes are the most chaotic objects. But the integrability of probes on black hole backgrounds is not in tension with the black hole chaos -- it is a consequence thereof. The horizon is a coarse-grained, thermodynamic description in which the nonintegrable, microscopic degrees of freedom have been integrated out. What remains is a highly symmetric effective geometry, and the only imprint of the underlying chaos on a probe is dissipation -- irreversible absorption and the exponential decay of quasinormal ringing. This is the familiar situation of a chaotic gas of hard spheres whose hydrodynamic description is smooth and simple, with microscopic chaos showing up only through viscosity.\footnote{We have explicitly demonstrated this mechanism on the example of highly supersymmetric, highly controlled LLM geometries in \cite{Berenstein:2025ese} -- only that the price to pay for the highly controlled calculation is that no black hole is formed upon coarse graining, but a naked singularity, i.e., a grayscale geometry.}

A microstate geometry is supposed to represent an object that undoes the coarse-graining. The horizon is replaced by a cap, absorption by reflection, quasinormal modes by normal modes, and a discrete spectrum is restored. At this point the question of probe chaos comes back. The integrable probe dynamics of the black hole and the chaotic probe dynamics of its microstate are thus two faces of the same statement, one for the ensemble average and one for an individual member.

\appendix

\section{Separability analysis of the Hamilton-Jacobi equation}

\subsection{$Q_1=Q_2$ Case}\label{secappa}

We start the separability analysis of the null geodesic motion in metric (\ref{eq:metsty2}) by inserting the separation ansatz (\ref{eq:sepAnsatz}) into the Hamilton-Jacobi equation (\ref{eq:hjgen}). Given that $\partial_u,\partial_v, \partial_{\varphi_1}, \partial_{\varphi_2}$ are all orthogonal to $\partial_r, \partial_\theta$ with $g_{r\theta}=0$, the metric (\ref{eq:metsty2}) is block diagonal. The $(r,\theta)$ block has $g_{rr}=\Sigma\sqrt{\mathcal P}/(r^2+a^2)$ and $g_{\theta\theta}=\Sigma\sqrt{\mathcal P}$ with no further $(u,v, \varphi_1,\varphi_2)$ admixture. Taking this into account, the Hamilton-Jacobi equation becomes (here we reprint the equation from the main text for convenience)
\be\label{eq:hj}
\left(r^2+a^2\right)\left(\partial_rS_r\right)^2+\left(\partial_\theta S_\theta\right)^2=\Sigma\mathcal W,
\ee
where
\be \label{eq:effpot}
    \Sigma \mathcal W = - \sqrt{\mathcal P} g^{AB}p_A p_B, \quad A,B \in \{u,v,\varphi_1,\varphi_2 \},
\ee
is quadratic in the momenta $p_A$ (to be defined in a while) and carries all of the mixed ($r,\theta$) dependence coming from the $(u,v,\varphi_1,\varphi_2)$ block. This is more conveniently written in the frame $\widetilde v = dv+\beta,~\widetilde u = du+\omega$:
\be \label{eq:met4}
    g_{AB} = \left[\begin{array}{cc|cc}
            0 & -1/\sqrt{\mathcal P} & \ {} & \ {}\\
            -1/\sqrt{\mathcal P} & -\mathcal F/\sqrt{\mathcal P} & \ {} & \ {}\\ \hline
            \ {} & \ {} & \sqrt{\mathcal P}(r^2+a^2) \sin^2 \theta & 0\\
            \ {} & \ {} & 0 & \sqrt{\mathcal P}r^2 \cos^2 \theta
\end{array}\right].
\ee
The important thing to notice is that the new dual frames are now $\{ \partial_u,~\partial_v,~\partial_{\varphi_1} - \omega_1 \partial_u - \beta_1 \partial_v,~\partial_{\varphi_2} - \omega_2 \partial_u - \beta_2 \partial_v\}$, and the associated conserved charges in this frame are
\be
    p_A = \{ p_u,~p_v,~M_i \equiv m_i - \beta_i p_v - \omega_i p_u \}, \quad i \in \{1,2\}.
\ee
After taking the inverse of the block diagonal metric (\ref{eq:met4}), we have
\be
    \mathcal W = \mathcal P \left(2 p_u p_v - \mathcal F p_u^2\right) - \frac{M_1^2}{(r^2 +a^2)\sin^2 \theta} - \frac{M_2^2}{r^2 \cos^2 \theta}.
\ee
All that said and done, the question of separability of the Hamilton-Jacobi equation (\ref{eq:hj}) now boils down to the question whether $\Sigma \mathcal W$ can be separated or not, i.e. written in the form
\be
\Sigma \mathcal W = \mathcal R(r) +  \Theta(\theta).\label{eq:sigmasep}
\ee
For closely related single-mode superstrata, \cite{Bena:2017upb} have found precisely this kind of structure directly: the $(1,0,n)$ family has a genuine conformal Killing tensor and is completely integrable, while the $(2,1,n)$ family only has it on a restricted locus. Let us now see if the form (\ref{eq:sigmasep}) can also be found for our case.

As a sanity check, let us first examine the momentum-free limit $b_4 = 0$ of (\ref{eq:metsty2}), i.e. the ordinary round, asymptotically flat, $k$-wound 2-charge D1-D5 supertube. We find, by direct construction of an explicit separated solution, that the null Hamilton-Jacobi equation separates for fully generic conserved momenta -- not merely on a restricted locus. This is in agreement with \cite{Lunin:2001dt} -- the round supertube is integrable. One can check that $\partial_r\partial_\theta\left(\Sigma \mathcal W\right) =0$ holds identically when tested on the complete, unexpanded expression for $\Sigma\mathcal W$ -- confirming that the split is not an artifact of how the terms happen to be grouped. As a further check, the $a \to 0$ (flat space) limit of (\ref{eq:om1})-(\ref{eq:beta}) collapses to the manifestly separated form
\be
    -\Sigma \mathcal W\big |_{a \to 0} = -\frac{2 p_up_v Q^2}{r^2}+ \frac{m_1^2}{\sin^2 \theta} + \frac{m_2^2}{\cos^2 \theta} \to \mathrm{const.},
\ee
reproducing the expected geodesic constants of motion for flat $\mathbb R^4$ in bipolar coordinates. We conclude that the equation (\ref{eq:hj}) in this simple, 2-charge case separates into a pair of first-order ODEs
\be \label{eq:sepODEs}
\left(r^2+a^2\right)\left(\partial_rS_r\right)^2 + \mathcal R = - \mathcal K, \quad \left(\partial_\theta S_\theta\right)^2 + \Theta = \mathcal K,
\ee
where $R$ and $\Theta$ are defined in eq.~(\ref{eq:sigmasep}) and the separation constant is $\mathcal K = p_\theta^2 + m_1^2/\sin^2\theta +m_2^2/\cos^2 \theta$ -- precisely the extra conserved quantity anticipated in the main text. Therefore, the sanity check works.

We will now consider more general cases $b_4 \neq 0$, where our real interest lies. Free data are $Q (\equiv Q_1=Q_2)$, $R$, and integers $k$, $n$, and $b_4$. Writing $\lambda = 2 |b_4|/Q$, the regularity condition becomes
\be \label{eq:reg}
a^2=\frac{Q^2-4|b_4|^2}{8k^2R^2}=\frac{Q^2(1-\lambda^2)}{8k^2R^2}.
\ee
We further introduce
\be
\nu\equiv\frac{1}{1-\lambda^2},~~\mu\equiv\frac{\lambda^2}{1-\lambda^2}=\lambda^2\nu,
\ee
so that (\ref{eq:reg}) and the definitions of $\mu$ and $\nu$ give
\be\label{eq:reg2}
\frac{Q^2}{a^2}=8k^2R^2\nu, \quad \frac{b_4^2}{a^2}=2k^2R^2\mu, \quad \frac{b_4^2}{k^2R^2a^2}=2\mu.
\ee
Now that the time has come to analyze in detail the structure of $\Sigma\mathcal{W}$ we have to specify the form of all metric functions. As we have explained in the main text, we are mainly interested in the decoupling limit, with AdS${}_3$ asymptotics.\footnote{For the $b_4=0$ (round supertube) case this was inessential; that solution is simple enough to analyze in its full form, and the integrability in a subspace of the whole space then follows by default. But for the general case we do need to avoid any inessential complications, so we immediately take the decoupling limit.} The metric functions then acquire the form:
\bea
&&\delta\equiv\Delta^{2kn}=\left(\frac{a^2\cos^2\theta}{\left(r^2+a^2\right)}\right)^{kn}\label{eq:Delta}\\
&&\mathcal P = \frac{Q^2-4b_4^2\delta}{\Sigma^2}, \quad \mathcal F = - \frac{b_4^2(1-\delta)}{k^2 R^2\Lambda}\\
&&\omega_1 = \frac{2\sin^2\theta}{\Sigma} \left[k^2R^2a^2+\frac{b_4^2}{kR}\frac{(r^2+a^2)(1-\delta)}{\Lambda} \right]\\
&&\omega_2=\frac{2\cos^2\theta}{\Sigma}\left[k^2R^2a^2+\frac{b_4^2}{kR}\frac{r^2(1-\delta)}{\Lambda}\right]\\
&&\beta_1=\frac{2kRa^2\sin^2\theta}{\Sigma}, \quad \beta_2 = - \frac{2kRa^2\cos^2\theta}{\Sigma}.
\eea
Next we introduce rescaled variables by putting
\be
x\equiv\frac{r}{a},~~u\equiv\cos^2\theta,
\ee
so that
\be
\Sigma=a^2\left(x^2+u\right),~~\Lambda=a^2\left(x^2+1-u\right),~~\delta=\left(\frac{u}{x^2+1}\right)^{kn}.
\ee
We further define
\be \label{eq:Pi}
    \Pi \equiv \frac{1-\delta}{x^2+1-u} = \frac{a^2\left(1 - \Delta^{2kn}\right)}{\Lambda},
\ee
so that, using (\ref{eq:reg2}), we get
\bea
    &&\mathcal F = - 2 \mu \Pi, \quad \beta_1 = \frac{2 k R (1-u)}{x^2 +u}, \quad \beta_2 = - \frac{2kR u}{x^2 + u}\label{eq:fbeta}\\
    &&\omega_1 = \beta_1 \left( 1 + 2 \mu (x^2+1)\Pi \right), \quad \omega_2 = - \beta_2 \left( 1 + 2 \mu x^2 \Pi \right). \label{eq:ombeta}
\eea
Finally, the effective potential (\ref{eq:effpot}) acquires the form
\be
\label{eq:pot}\Sigma\mathcal W =\Sigma\left[\mathcal P \left(2p_up_v-\mathcal F p_u^2\right)-\frac{M_1^2}{(r^2+a^2)\sin^2\theta}-\frac{M_2^2}{r^2\cos^2\theta}\right]\equiv\Sigma\mathcal P \left(2p_up_v-\mathcal F p_u^2\right)+\widetilde{W},\\
\ee
where $\widetilde{W}$ encapsulates the last two factors in $\Sigma\mathcal{W}$, i.e., the terms proportional to $M_1^2$ and $M_2^2$. The factors $M_i$ are found from (\ref{eq:ombeta}), introducing $p_u+p_v\equiv 2\mathfrak p$ and $p_v-p_u\equiv\mathfrak q$,
\bea
M_1&=&m_1-2\mathfrak p \beta_1 - 2\mu\beta_1(x^2+1)\Pi p_u = M_1^{(0)}-A_1, \quad M_1^{(0)} \equiv m_1-2\mathfrak p \beta_1\label{eq:mnew1}\\
M_2&=&m_2-\mathfrak q \beta_2-\left(-2\mu\beta_2 x^2\Pi p_u\right)~ = M_2^{(0)}-A_2, \quad M_2^{(0)}\equiv m_2-\mathfrak q \beta_2\label{eq:mnew2}.
\eea
Now concentrate on the contribution $\widetilde{W}$ in (\ref{eq:pot}). In the rescaled variables, it becomes
\be
\widetilde{W}=-\frac{\left(x^2+u\right)M_1^2}{\left(x^2+1\right)(1-u)}-\frac{\left(x^2+u\right)M_2^2}{x^2u}.
\ee
Writing $M_{1,2}$ as $M_{1,2}^{(0)}-A_{1,2}$ in accordance with (\ref{eq:mnew1}-\ref{eq:mnew2}), we get quadratic terms in $A_{1,2}$, denoted by $\widetilde{W}_2$, linear terms in $A_{1,2}$, denoted by $\widetilde{W}_1$ and terms independent of $A_{1,2}$. The terms linear in $A_{1,2}$ have prefactors that collapse to a single term:
\be
\widetilde{W}_1=-2\frac{\left(x^2+u\right)A_1}{\left(x^2+1\right)(1-u)}-2\frac{\left(x^2+u\right)A_2}{x^2u}=16kR\mu p_u\Pi,
\ee
The quadratic terms give
\be
\widetilde{W}_2=-\frac{\left(x^2+u\right)A_1^2}{\left(x^2+1\right)(1-u)}-\frac{\left(x^2+u\right)A_2^2}{x^2 u} =-\frac{16k^2R^2\mu^2 p_u^2\Pi^2}{x^2+u}\left[(1-u)(x^2+1)+ux^2\right].\label{eq:quadw}
\ee
Now using the simple identity $(1-u)(x^2+1)+ux^2=x^2+1-u$ and $\Pi (x^2+1-u)=1-\delta$ (see eq.~(\ref{eq:Pi})), the quadratic term (\ref{eq:quadw}) becomes
\be \label{eq:muquad}
\widetilde{W}_2=-\frac{16k^2R^2\mu^2 p_u^2\Pi(1-\delta)}{x^2+u}.
\ee
The twist is that this term gets cancelled by a contribution from the second term of the potential (\ref{eq:pot}), the one proportional to $p_u^2$:
\be
-\Sigma\mathcal P \mathcal F p_u^2 = \frac{16k^2R^2\mu p_u^2\Pi}{x^2+u}+\frac{16k^2R^2\mu^2 p_u^2\Pi(1-\delta)}{x^2+u}.
\ee
The second piece of this term cancels the quadratic term (\ref{eq:muquad}). Collecting only linear terms in $\mu$ we get
\be
\frac{16 k^2 R^2 \mu p_u p_v (1-\delta)}{x^2+u} + \frac{16 k^2 R^2 \mu p_u^2\Pi}{x^2 + u} = 8 k R \mu p_u \Pi (M_1^{0} + M_2^{0}).
\ee
Finally, collecting all the terms of $\Sigma\mathcal{W}$ and taking into account the cancellation that we found above, we get (for convenience we reproduce here the expression given in the main text):
\be
    \Sigma \mathcal W=\Sigma\mathcal W \big |_{b=0}-\frac{1}{2}\mathcal C \mathcal F, \quad \mathcal C \equiv 8kRp_u\left(m_1+m_2+2kRp_v\right),\label{eq:potfinalapp}
\ee
where $\mathcal{F}$ is some function of all variables in the problem. This expression is easy to analyze for separability: we know that the first term, which reduces to the round supertube case, is separable, and the second, in general, is not (because $\mathcal{F}$ in general depends on all variables). Therefore, separability is only achieved if $\mathcal{C}$ vanishes, or if $\mathcal{F}$ simplifies sufficiently to become separable. This is the basis of the analysis of (\ref{eq:potfinal}) performed in the main text.

\subsection{$Q_1\neq Q_2$ Case}\label{secappb}

Inspecting the metric specified by (\ref{eq:qgenomegar}-\ref{eq:Pgen}), we find that the surviving isometries are $\partial_u,~\partial_{\varphi_1}$ and $2kR \partial_v + \partial_{\varphi_2}$. For the Hamilton-Jacobi function we take the ansatz (\ref{eq:ansatz2}): $S=p_uu+p_vv+m_1\varphi_1+m_2\varphi_2+S_r+S_{\theta}+S_\zeta$. Then $\partial_v S$, $\partial_{\varphi_2}S$ and $\partial_\zeta S$ act as $\partial_v \to i p_v + \frac{1}{2R}\partial_\zeta$, $\partial_{\varphi_2}\to i m_2 - k\partial_\zeta$ and $\partial_\zeta\to m$, so that the mode $m$ carries shifted charges:
\be \label{eq:shftch}
    p_v^{(m)} = p_v + \frac{m}{2R}, \quad m_2^{(m)}= m_2 - mk.
\ee
Only three combinations are conserved: $p_u$, $m_1$, and
\be
    2 kR p_v^{(m)} + m_2^{(m)} = 2kR p_v + m_2,
\ee
the charge conjugate to the Killing $2kR \partial_v + \partial_{\varphi_2}$ (which is independent of $m$). Neither $p_v$ nor $m_2$ is separately conserved. Our strategy now is to eliminate the fake degrees of freedom, corresponding to integrals of motion (which are not manifest in the above formulation but will become so after some coordinate frame transformations), and obtain a lower-dimensional geodesic action which is easier to analyze.

In the frame adapted to the fibration the metric is block diagonal
\be
    g = - \frac{1}{\sqrt{\mathcal P}} \left( \sigma \otimes \mu + \mu \otimes \sigma \right) - \frac{\mathcal F}{\sqrt{\mathcal P}} \sigma \otimes \sigma + \sqrt{\mathcal P} h, \quad \sigma = dv+\beta, \quad \mu = du + \omega,
\ee
with $h$ the flat base metric, carrying legs only along $x^m = (r,\theta,\varphi_1,\varphi_2)$. Take the frame dual to the coframe $\{ \sigma, \mu, dx^m \}$:
\be
    E_v = \partial_v, \quad E_u = \partial_u, \quad E_m = \partial_m - \beta_m \partial_v - \omega_m \partial_u,
\ee
for which indeed $\sigma(E_m) = \beta_m - \beta_m = 0$ and $\mu(E_m) = \omega_m - \omega_m =0$. Since $h$ has no $u,~v$ legs, every mixed component of $g$ in this frame collapses:
\bea
    g(E_v,E_v) &=& - \mathcal F/\sqrt{\mathcal P}, \\
    g(E_v, E_u) &=& - 1/\sqrt{\mathcal P}, \\
    g(E_u,E_u) &=& g(E_{u,v},E_m)= 0, \\
    g(E_m,E_n) &=& \sqrt{\mathcal P} h_{mn}.
\eea
The $2\times 2$ null block has $\mathrm{det} = - 1/\mathcal P$ and its inverse is
\be
    \begin{pmatrix}
        0 & -\sqrt{\mathcal P} \\
        -\sqrt{\mathcal P} & \sqrt{\mathcal P}\mathcal F
    \end{pmatrix}
\ee
while the base block gives $h^{mn}/\sqrt{\mathcal P}$.

The frame components of $p_M = \partial_M S$ are $p(E_v)=p_v$, $p(E_u)=p_u$ and
\be \label{eq:ch}
    \mathcal P_m \equiv p(E_m) = p_m - \beta_m p_v - \omega_m p_u.
\ee
Since $\beta$ has legs only along $\varphi_1$, $\varphi_2$, whereas $\omega$ also has $dr$ and $d\theta$ legs, (\ref{eq:ch}) produces both the familiar angular charges and the minimal coupling:
\be
    \mathcal P_r = p_r - \omega_r p_u, \quad \mathcal P_\theta = p_\theta - \omega_\theta p_u, \quad \mathcal P_{\varphi_1} = M_1, \quad \mathcal P_{\varphi_2} = M_2.
\ee
Assembling everything together, with $h^{mn}= \mathrm{diag}\left( \frac{r^2+a^2}{\Sigma}, \frac{1}{\Sigma}, \frac{1}{(r^2+a^2)\sin^2\theta}, \frac{1}{r^2 \cos^2 \theta} \right)$, one gets:
\be
    S=g^{MN}p_Mp_N = - 2 \sqrt{\mathcal P} p_u p_v + \sqrt{\mathcal P} \mathcal F p_u^2 + \frac{1}{\sqrt{\mathcal P}} h^{mn} \mathcal P_m \mathcal P_n.
\ee
Finally, multiplying by the positive factor $\Sigma\sqrt{\mathcal P}$ gives the action for the null geodesic on the supergravity solution:
\bea
    &&S=\Sigma \sqrt{\mathcal P} g^{MN}p_M p_N = (r^2+a^2) (p_r-\omega_r p_u)^2 + (p_\theta - \omega_\theta p_u)^2 - \Sigma \mathcal W =0 \\
    &&\mathcal W = \mathcal P \left( 2p_u p_v^{(m)} - \mathcal F p_u^2 \right) - \frac{M_1^2}{(r^2+a^2)\sin^2 \theta}-\frac{M_2^2}{r^2 \cos^2 \theta} \label{eq:ham2}\\
    &&M_i = m_i - \beta_i p_v^{(m)} - \omega_i p_u. \label{eq:Ms}
\eea
Expanding $\Sigma \mathcal W$ in the $\zeta$-momentum $m$ as $\Sigma \mathcal W=A_2m^2+A_1m+A_0$, one finds all coefficients in closed form
\bea
A_2 &=& -\frac{k^2 \Lambda}{\left(r^2+a^2\right) \cos^2 \theta} = \frac{k^2 a^2}{r^2+a^2}- \frac{k^2}{\cos^2 \theta}\label{eq:a2}\\
A_1 &=& \frac{\Sigma \mathcal P p_u}{R}+ \frac{2ka^2 \mathcal M_1}{r^2+a^2} + \frac{2k \mathcal M_2}{\cos^2\theta}\label{eq:a1}\\
A_0 &=& \Sigma \left[ \mathcal P \left( 2 p_u p_v - \mathcal F p_u^2\right) - \frac{\mathcal M_1^2}{\left( r^2 +a^2\right)\sin^2 \theta} - \frac{\mathcal M_2^2}{r^2 \cos^2 \theta} \right]\\
&&\mathcal M_1 \equiv M_1 + \frac{ka^2 \sin^2 \theta}{\Sigma}m,~~\mathcal M_2 \equiv M_2 + \frac{kr^2}{\Sigma}m.\label{eq:a0}
\eea
As we see, the quadratic term is separable. The null geodesics are consequently the orbits of the 3-degree-of-freedom Hamiltonian (which is stated also in the main text in eq.~(\ref{eq:hamcoh3}) and which we copy here for convenience):
\be \label{eq:hamcoh3app}
    \bar{H} = \left( r^2+a^2 \right) \left( p_r-\omega_r p_u \right)^2 + \left( p_\theta - \omega_\theta p_u \right)^2 - A_2 m^2 - A_1m - A_0 = 0. 
\ee
This Hamiltonian is considerably simpler than the original full, higher-dimensional motion. Its analysis is not long and is given in the main text.

Finally, let us perform a consistency check: we will show that setting $Q_1 = Q_2$ reduces to the cohomogeneity-2 problem. This directly implies $\chi =1$ and $b_1 = - b_2$. Next we have $\mathcal C_1 =0$, implying that $\omega$ loses its $r$ and $\theta$ legs, thus the left-hand sides of (\ref{eq:hj}) and (\ref{eq:hjcoh3}) are the same. Furthermore, this implies (\ref{eq:om1gen}) $\to$ (\ref{eq:om1}), (\ref{eq:om2gen}) $\to$ (\ref{eq:om2}), and (\ref{eq:Pgen}) $\to$ (\ref{eq:P}), so all actors play the same roles in this setup. Notice that $\mathcal F$ and $\beta$ were never phase-dependent, so this leaves $\bar H$ independent of $\zeta$, thus $\dot m = - \partial_\zeta \bar H =0$ and $m$ is conserved. With $m$ frozen, the shifted charges (\ref{eq:shftch}) are constants. Relabelling the shifted charges back to $p_v$ and $m_2$, we see that the coefficients are $\zeta$-free and
\be
    \Sigma \mathcal W = \Sigma \left[ \mathcal P \left( 2 p_u p_v - \mathcal F p_u^2 \right) - \frac{M_1^2}{\left( r^2+a^2 \right) \sin^2 \theta}  - \frac{M_2^2}{r^2 \cos^2 \theta} \right], \quad M_i = m_i - \beta_i p_v - \omega_i p_u,
\ee
which is precisely the cohomogeneity-2 potential. Note that $m$ is not forced to vanish: each value of $m$ gives the cohomogeneity-2 problem at the shifted charges $p_v + m/2R, ~m_2-mk$, so $Q_1=Q_2$ yields a one-parameter family.

\section{Numerical method}\label{secappbb}

\subsection{Cohomogeneity-2 solutions}

We start from a simple observation: from (\ref{eq:ww}), the reduced potential $\Sigma \mathcal W$ on the slice $p_u=p_v\equiv\epsilon$ is exactly quadratic in $\epsilon$, so the massless wave equation becomes a quadratic eigenvalue problem, schematically of the form
\be
\left(L_0+\epsilon L_1+\epsilon^2L_2\right)f=0.
\ee
This is an advantage over a full nonlinear root search. We discretize the equation by pseudospectral collocation on Chebyshev-Gauss nodes in the regular variables $x = r^2/\left( r^2 +a^2 \right)$ and $w = \cos \left(2 \theta \right)$, in which the kinetic operators take the polynomial-coefficient form
\be \label{eq:regop}
    4x \left(1-x\right)^2 \partial_x^2 + 4(1-x)^2 \partial_x \quad \mathrm{and} \quad 4 \left(1 - w^2 \right) \partial_w^2 - 8 w \partial_w.
\ee
For statistical analysis each mode is mapped to coordinates in which the 2-dimensional problem is a genuine flat-space Helmholtz equation as in (\ref{eq:helmholtz}). Note that the eigenvalue problem is solved in collocation (Chebyshev-Gauss) variables -- the flat-frame variables are used only to express the computed modes in a form to which random-wave statistics apply.

\subsection{Cohomogeneity-3 solutions}

Recall that the general Style-1 superstratum depends, in addition to $r$ and $\theta$, on the fractionated phase $\zeta = v/2R - k\varphi_2$ through the momentum-carrying wave $\propto b_4 e^{i n \zeta}$ of the first-layer functions $Z_{1,2,4}$. When $Q_1=Q_2$ the coiffuring via Style 1 cancels this dependence from the metric yielding cohomogeneity-2; for $Q_1 \neq Q_2$ it survives, the isometry is reduced to the three commuting Killing vectors, and the massless wave equation no longer separates into a single 2-dimensional problem. Our method rests on an exact property of this background which makes the loss of the fourth Killing vector tractable.

The background functions are assembled from the layer solutions (\ref{eq:qgenomegar})-(\ref{eq:Fgen}) and (\ref{eq:Pgen}) in the decoupling limit. The blocks $A_0,~A_\pm$ are the $\zeta$-Fourier coefficients of the metric, obtained by a fast Fourier transform over $N_\zeta$ points. In $(r, \theta)$ we use pseudospectral collocation on Chebyshev-Gauss nodes in the regular variables $x = r^2/\left( r^2 + a^2 \right) \in (0,1)$ and $w = \cos 2\theta \in (-1,1)$, in which the kinetic operators take the same form as in eq.~(\ref{eq:regop}). The coordinate singularities at the cap and at the poles are mapped to interval endpoints that the Chebyshev-Gauss nodes never touch, and the Frobenius behavior $f \sim r^{|m_2|},~|\theta|^{|m_1|}$ becomes polynomial in $(x,w)$. The regular branch is therefore selected automatically, no boundary conditions are imposed as matrix rows, and the potential is bounded on the grid, so that machine precision suffices. Each block (\ref{eq:masKGcoh3}) is formed by contracting the appropriate $D^{(m)}$ vectors with the discretized $A$-blocks.

On the slice $p_u=p_v = \tilde \epsilon$ every matrix element of (\ref{eq:masKGcoh3}) is at most quadratic in $\tilde\epsilon$, so that the truncated tower $m \in \{-M n, \cdots, Mn \}$ becomes a sparse, block-tridiagonal quadratic eigenvalue problem $\left(L_0 + \tilde \epsilon L_1 + \tilde \epsilon^2 L_2 \right) f =0$ of dimension $(2M+1)N_xN_w$. We solve it by linearization and shift-invert Arnoldi iterations about a target $\tilde \epsilon_\sigma$, discarding those roots of the linearization that fail the residual test $|L(\tilde \epsilon)| < 10^{-6}|L\cdot||f|$ or have a non-negligible imaginary part. An eigenvector is the stacked set of block fields $(f_{-Mn}, \cdots f_{Mn})$.

\section{Statistical tests of the random wave behavior and separability}\label{secappc}

Here we explain in detail how we probe wave chaos and which numerical procedures and statistical tests we use. 

The solver returns, for each mode, the eigenfrequency and the nodal values of the reduced field $f(x,w)$ on the Chebyshev-Gauss grid in $(x,w)$ coordinates. Diagnostics are not evaluated on this grid, whose nodes cluster in the boundaries, but on a uniform grid in flat coordinates $(\xi, \theta)$, in which the reduced wave equation takes the Helmholtz form (\ref{eq:helmholtz}).

The practical procedure is thus the following. Each field is brought to the frame in which the 2-dimensional problem is a Helmholtz equation (see the discussion around eq.~(\ref{eq:helmholtz})), evaluated at the shifted charges, and the physical field on a $\zeta$-slice is $\psi = \Re \sum_m \psi_m e^{im\zeta}$. On these fields we evaluate the diagnostics used throughout: the Kolmogorov-Smirnov distance between the actual 2-point correlator and the Berry prediction $J_0(s)$, the Kolmogorov-Smirnov distance between the actual intensity distribution and the Porter-Thomas law, and rank-1 fraction.

The 2-point correlator is computed by tabulating one mode over the lattice, tabulating a copy of itself separated by some displacement (allowed by the lattice), tabulating their product over all allowed displacements, and averaging the product over all displacements with equal norm. The result is expressed as a function of locally rescaled separation:
\be \label{eq:s}
s\equiv \frac{1}{2}\left(k_{\mathrm{eff}}\left(\vec x_1\right)+k_{\mathrm{eff}}\left(\vec{x}_2\right)\right)\cdot\left|\vec{x}_1-\vec{x}_2\right|,
\ee
so that Berry's prediction from eq.~(\ref{eq:berry2}) carries no free parameter. Therefore, we do not perform any fit but just compare the random-wave prediction and the data.

An additional subtlety is that our effective potential is nontrivial and the values of $k_\mathrm{eff}$ in different directions differ. Direct comparison with eq.~(\ref{eq:berry1}), which only depends on separation, thus requires us to perform averaging over the angles. We take the separation vector $\vec \Delta\equiv\vec{x}_1-\vec{x}_2$ from a lattice of displacements in the $(\xi,\theta)$ plane that covers all directions and all lengths up to three wavelengths ($s \leq 6 \pi$). Pairs at the same $s$ but from different directions are then averaged over.

The intensity statistics are built per mode from the same flat-frame ($\xi, \theta$) field $\psi$ on the same uniform grid. The intensity is taken at every grid point of the allowed region $I_j = \psi_j^2, ~j \in \mathcal M$, and divided by its arithmetic mean over those points,
\be
    I_j \to \frac{\psi_j^2}{ N_{\mathcal M}^{-1} \sum_{i \in \mathcal M} \psi_i^2},
\ee
so that $\langle I \rangle =1$ for every mode by construction. The mean is the flat frame area integral over $d \xi d\theta$, every grid cell counting equally; this is the measure in which $\psi$ is a Helmholtz field (\ref{eq:helmholtz}), and hence the one in which the Berry conjecture predicts Gaussian amplitudes. Points outside $\mathcal M$ never enter. Comparison with the Porter-Thomas law is performed again through the Kolmogorov-Smirnov distance of the empirical cumulative distribution from the one obtained from the Porter-Thomas law (\ref{eq:PTlaw}):
\be
    \mathsf{KS} = \max_I \left| F(I) - F_{PT}(I) \right|, \quad F(I) = \int_0^I P(I')dI', \quad F_{PT} = \mathrm{Erf}\left(\sqrt{\frac{I}{2}}\right).\label{eq:ksdistance}
\ee
This measure is free of binning and smoothing choices. The finite-sample floor ($\simeq 0.007$ at our grid sizes) is calibrated on synthetic random waves; a mode with $\mathsf{KS} \lesssim 0.02$ is indistinguishable from a random wave at this resolution. We regard the measure (\ref{eq:ksdistance}) as similar in spirit but improved in terms of stability in comparison to the inverse-square distance $\chi^{-2}$ that we have used to judge the proximity of 2-point correlations to the Berry expression in \cite{Djukic:2026ejk}.

Finally, separability is estimated in the following way. For an eigenmode $\psi$ sampled on the flat-frame grid $(\xi_i,\theta_j)$ we form the matrix $\Psi_{ij}= \psi(\xi_i,\theta_j)$ and take its singular value decomposition 
\be
    \Psi_{ij} = \sum_k s_k u_k(\xi_i) v_k(\theta_j),~\text{with}~s_1\geq s_2 \geq \cdots.
\ee
The rank-1 fraction is the fraction of the mode's norm carried by its best single product $R(\xi)\Theta(\theta)$:
\be \label{eq:r1}
    \mathcal R_1 = \frac{s_1^2}{\sum_k s_k^2}.
\ee
It equals unity on a separable wavefunction, and takes values $\simeq 0.2-0.35$ for a Gaussian random wave.

\acknowledgments

We are grateful to Iosif Bena, David Berenstein, Nikolay Bobev, Sean Colin-Ellerin, Rapha\" el Dulac, Siyul Lee, Yixuan Li, Kyriakos Papadodimas, Milica Stepanovi\'c, Marija Toma\v sevi\' c and Erez Urbach for stimulating discussions. V.~\DJ. is grateful to Institute for Theoretical Physics at KU Leuven, Institut de Physique Théorique (IPhT), CEA Saclay, and CERN Department of Theoretical Physics for their hospitality and for the opportunity to present preliminary results of this work. We acknowledge support by the Delta ITP consortium, a program of the Netherlands Organization for Scientific Research (NWO) funded by the Dutch Ministry of Education, Culture and Science (OCW). Work at the Institute of Physics Belgrade is funded by the Ministry of Science, Technological Development and Innovation and by the Science Fund of the Republic of Serbia. The work on Sections 2-4 was supported by Russian Science Foundation Grant No. 24-72-10061 [https://rscf.ru/project/24-72-10061/] and performed at Steklov Mathematical Institute of Russian Academy of Sciences (M.~\v{C}.).

This research has made use of large-language models Claude Opus 5 and Fable 5 (Anthropic) with parameter space scans, visualizations, and the development of some of the numerical code underlying this work. All codes have been tested and benchmarked by the authors. The authors take full responsibility for the content of the paper.

\bibliography{LLM.bib}

\end{document}